\documentclass[usenatbib]{mnras}

\usepackage{graphicx,todonotes}
\usepackage{latexsym}
\usepackage{natbib,amssymb}
\usepackage{breakurl}
\usepackage{amsmath}
\usepackage[OT2,T1]{fontenc}
\usepackage{comment}
\graphicspath{{figures/}}

\newcommand{\rev}[1]{#1}

\newcommand{\beq}{\begin{equation}}
\newcommand{\eeq}{\end{equation}}

\newcommand{\galsim}{\texttt{GalSim}}
\newcommand{\sersic}{{S\'ersic}}
\newcommand{\synpipe}{\texttt{SynPipe}}

\title[Weak lensing simulations]{Weak lensing shear calibration with simulations of the HSC survey}

\author[Mandelbaum et~al.]{Rachel Mandelbaum$^1$\thanks{\tt rmandelb@andrew.cmu.edu},
Fran\c{c}ois Lanusse$^1$,
Alexie Leauthaud$^2$,
Robert Armstrong$^3$,
\newauthor
Melanie Simet$^{4,5}$,
Hironao Miyatake$^{4,6,7,8}$,
Joshua E.\ Meyers$^{9,3}$,
\newauthor
James Bosch$^3$,
Ryoma Murata$^6$,
Satoshi Miyazaki$^{10}$,
Masayuki Tanaka$^{10}$
\\$^1$McWilliams Center for Cosmology, Department of Physics, Carnegie Mellon University,
  Pittsburgh, PA 15213, USA
\\$^2$Department of Astronomy and Astrophysics, University of California, Santa Cruz,
  1156 High Street, Santa Cruz, CA 95064 USA
\\$^3$Department of Astrophysical Sciences, 4 Ivy Lane, Princeton University, Princeton,
  NJ 08544
\\$^4$Jet Propulsion Laboratory, California Institute of Technology, Pasadena, CA 91109, USA
\\$^5$University of California, Riverside, 900 University Avenue, Riverside, CA 92521,
  USA
\\$^6$Kavli Institute for the Physics and Mathematics of the Universe (Kavli IPMU, WPI),
UTIAS,
  \\Tokyo Institutes for Advanced Study, The University of Tokyo, Chiba 277-8583, Japan
\\$^7$Institute for Advanced Research, Nagoya University, Nagoya 464-8601, Aichi, Japan
\\$^8$Division of Physics and Astrophysical Science, Graduate School of Science, Nagoya University, Nagoya 464-8602, Aichi, Japan
\\$^9$Kavli Institute for Particle Astrophysics and Cosmology, Department of Physics, \\Stanford
University, Stanford, CA 94305, United States of America
\\$^{10}$National Astronomical Observatory of Japan, Mitaka, Tokyo 181-8588, Japan
}

\begin{document}
\date{\today}
\maketitle

\begin{abstract}
  We present results from a set of simulations designed to constrain the weak lensing shear
  calibration for the Hyper Suprime-Cam (HSC) survey.  These simulations include HSC observing
  conditions and galaxy images from the Hubble Space Telescope ({\em HST}), with fully realistic
  galaxy morphologies and the impact of nearby galaxies included.  We find that the inclusion of
  nearby galaxies in the images is critical to reproducing the observed distributions of galaxy
  sizes and magnitudes, due to the non-negligible fraction of unrecognized blends in
  ground-based data, even with the excellent typical seeing of the HSC survey (0.58\arcsec\ in the
  $i$-band).  Using these simulations, we detect and remove the impact of selection biases due to
  the correlation of weights and the quantities used to define the sample (S/N and apparent size)
  with the lensing shear.  We quantify and remove galaxy property-dependent multiplicative and
  additive shear biases that are intrinsic to our shear estimation method, including
  a $\sim 10$ per cent-level multiplicative bias due to the impact of nearby galaxies and unrecognized
  blends.  Finally, we check the sensitivity of our shear calibration estimates to other cuts made on the simulated
  samples, and find that the changes in shear calibration are well within the requirements for HSC
  weak lensing analysis.  Overall, the simulations suggest that the weak lensing multiplicative
  biases in the first-year HSC shear catalog are controlled at the 1 per cent level.
\end{abstract}


\section{Introduction}

The $\Lambda$CDM cosmological model is dominated by dark matter and dark energy.  Weak gravitational
lensing is among the most promising ways of measuring the large-scale distribution of dark matter,
via the deflections of light due to intervening matter along the line-of-sight, which both magnifies
and distorts galaxy shapes \citep[for recent reviews,
see][]{2013PhR...530...87W,2015RPPh...78h6901K}.  This large-scale matter distribution in turn
reveals the growth of structure as a function of time, which can be used to
constrain the equation of state of dark energy 
\citep[e.g.,][]{2013MNRAS.432.2433H,2013MNRAS.432.1544M,2015ApJ...806....2M,2016ApJ...824...77J,2017MNRAS.465.1454H,2017arXiv170801538T}.

While weak lensing is a powerful cosmological measurement, it is also very technically challenging
due to the many sources of systematic errors that must be reduced below the statistical floor.
Among the ongoing wide-area sky surveys that are used for lensing measurements, the Hyper
Suprime-Cam (HSC, Miyazaki et~al.~{\em in prep}.; Komiyama et~al.~{\em in prep}.; Kawanomoto
et~al.~{\em in prep}.; Furusawa et~al.~{\em in prep.}) survey\footnote{\url{http://hsc.mtk.nao.ac.jp/}} \citep{2018PASJ...70S...4A,2018PASJ...70S...8A} has a unique combination of
depth and high-resolution imaging that gives it a longer redshift baseline and the ability to
measure a high source number density, 24.6 (raw) and
  21.8 (effective) arcmin$^{-2}$, even with relatively conservative cuts on the apparent magnitude
  of the galaxies used for the weak lensing analysis. In \citet{2018PASJ...70S..25M}, we described
empirical tests using the HSC galaxy shape catalog to demonstrate that shape measurement-related
systematic errors that can be determined empirically are well-controlled.  The focus of this
work is to demonstrate our methods of controlling for and removing multiplicative and additive bias in ensemble
shear estimates (the first of which cannot in general be detected using empirical null tests).

There are several sources of multiplicative bias in weak lensing measurements
that involve measuring and averaging galaxy shapes.  The presence of
noise in the galaxy images results in a bias in maximum-likelihood shape estimators
\citep[`noise bias';
e.g.,][]{2002AJ....123..583B,2004MNRAS.353..529H,2012MNRAS.427.2711K,2012MNRAS.424.2757M,2012MNRAS.425.1951R}. When estimating shears in a way that assumes a particular galaxy
model, the shears can be biased if the galaxy light profiles are not correctly described by that
model \citep[`model bias':][]{2010MNRAS.404..458V, 2010A&A...510A..75M}.  More generally, any method based
on the use of second moments to estimate shears cannot be completely independent of the details of
the galaxy light profiles, such as the overall galaxy morphology and presence of detailed
substructure \citep{2007MNRAS.380..229M,2010MNRAS.406.2793B,2011MNRAS.414.1047Z}.  Since these
issues are a generic feature of second moment-based methods that involve averaging galaxy shape
estimates, we must estimate their impact on shear estimation in the HSC survey.  There are also
selection biases due to the quantities used for selection \citep[e.g.,][]{2004MNRAS.353..529H,2016MNRAS.460.2245J} or for weighting the
ensemble lensing shear estimates \citep[e.g.,][]{2017MNRAS.467.1627F} depending on the galaxy shape, and hence
upweighting galaxies that are preferentially aligned with the shear and/or PSF anisotropy
directions.

Several methods have been proposed and developed for estimating and removing shear calibration
biases.  The weak lensing community has a long-standing tradition of using simulated images to test
shear estimation
\citep[e.g.,][]{2006MNRAS.368.1323H,2007MNRAS.376...13M,2010MNRAS.405.2044B,2012MNRAS.423.3163K,2013ApJS..205...12K,2015MNRAS.450.2963M}.
These have the advantage of enabling a test with a known `ground truth', and the disadvantage that
the results in general depend on the realism of the galaxy populations and other aspects of the
simulation.  The onus is thus on the simulator to ensure that the simulations are a faithful
representation of reality for any image characteristics that actually affect the fidelity of shear
estimation (including galaxy populations, PSF models, etc.).  Alternatively, a 
self-calibration method using the data itself has been proposed
\citep[Metacalibration:][]{2017arXiv170202600H,2017ApJ...841...24S}; this method involves
determining the response of a shear estimator to a lensing shear using re-simulated real data, which
by definition includes the correct galaxy population.  
CMB lensing has been identified as a means for calibrating relative biases due to shear and/or
photometric redshift error in optical shear estimates \citep[e.g.,][]{2017PhRvD..95l3512S}.
As shown there, a more powerful CMB experiment than currently exists would be needed to calibrate the shear to the precision needed
for HSC.  For HSC, CMB lensing can only serve as a nicely independent sanity check rather than a
means of estimating detailed shear calibration bias factors.

For the HSC survey first-year weak lensing science, we use simulations
to derive corrections for shear calibration biases.  The focus of
this paper is to describe the simulation process, illustrate the level of realism, and explain how they were
used to estimate and remove shear calibration biases and additive errors.  Our approach is to apply
corrections based on the simulations, and the uncertainties on those corrections then become a part of
our systematic error budget.  As described in \citet{2018PASJ...70S..25M},
for future HSC science releases we hope to have an integrated metacalibration routine that we can
use as our primary method of calibrating our ensemble shear estimates, with simulations used only as
a separate sanity check.  Moreover, a separate simulation approach,
\synpipe, which 
involves injecting galaxies into the real HSC data, was used to address a different set of problems
(\citealt{2018PASJ...70S...6H}, Murata et~al.~{\em in prep.}). The simulation
pipeline described in this work is optimized for estimation of shear calibration biases with a
larger ensemble of simulated galaxies. 

Finally, a completely different solution is to adopt a method of estimating ensemble
shear without measuring and averaging per-object shapes \citep[e.g.,][]{2016MNRAS.459.4467B}.  Adopting such
a method eliminates the need to estimate and remove noise and model bias. 
Armstrong et~al.~({\em in prep.}) will demonstrate the use of one such method for HSC, and compare 
the resulting ensemble shears with the catalog from
\citet{2018PASJ...70S..25M} after using the simulation-based calibration bias corrections described
in this paper.  Agreement between these two conceptually different approaches would further validate
the results in this work.

This paper is structured as follows. The simulation software and inputs such as galaxy populations
are described in Section~\ref{sec:software} and~\ref{sec:siminputs}, respectively.  The software
used to analyze the simulations and derive shear calibration biases is described in
Section~\ref{sec:analysis}, and results are in Section~\ref{sec:results}.  We conclude in
Section~\ref{sec:conclusions}.

\section{Software}\label{sec:software}

\subsection{Image simulations}

The open-source image simulation package
\galsim\footnote{\url{https://github.com/GalSim-developers/GalSim}} \citep{2015A&C....10..121R}
\texttt{v1.4.2} was used to produce the simulations described in this work.  The inputs to the simulation -- galaxy
models, PSFs, noise variances, and correlations -- are described in Section~\ref{sec:siminputs}.
Here we outline the way the simulations were designed overall.

As in the GREAT3 challenge \citep{2014ApJS..212....5M,2015MNRAS.450.2963M}, the simulations were
produced in terms of {\em subfields}, each with a grid of $100\times 100$ galaxy postage stamps (potentially with other nearby structures surrounding the $10^4$ objects of
interest, as will be described below). 
All galaxies within a subfield had the same lensing shear, PSF, and noise
level.  Each of the $800$ subfields had a different shear, PSF, and noise level.  Determination of shear
calibration biases and additive systematics was based on the ensemble shear estimates for each subfield.
To reduce statistical errors on the shear biases, galaxies were arranged such that the intrinsic
noise due to non-circular galaxy shapes (`shape noise') was nearly cancelled out by incorporating
pairs of galaxies rotated at $90^\circ$ with respect to each other \citep{2007MNRAS.376...13M}.

Our strategy is to produce simulations of coadded images, rather than simulating individual
exposures and then coadding them.  As a consequence of this choice, we cannot use these simulations
to investigate systematics in the production of coadds, such as relative astrometric errors.  In
addition, since we adopt a fixed sky level (zero) and use PSF models directly for shear estimation, our analysis of the
simulations does not include systematics due to sky level misestimation or PSF modeling errors.  
We do not insert star images, and hence there is no stellar contamination. 
The impact of astrometric errors, sky misestimation, PSF modeling errors, and star-galaxy contamination
on weak lensing in HSC were addressed using empirical tests in \citet{2018PASJ...70S..25M}.
Other effects that are not tested include wavelength-dependent systematics (e.g., chromatic PSFs),
instrumental and detector defects, artifacts, or non-linearities; non-weak shear signals; and
flexion. 

However, as described in Sec.~\ref{subsec:hsc-pipeline}, our HSC pipeline analysis does include object detection, deblending,
and selection in addition to shear estimation.  Since our input images include other objects near the
galaxies of interest on the grid, the deblending test is non-trivial.
%

We used the \galsim\ config interface to produce the simulations from yaml files. Images were
rendered using Fourier space rendering, which is the only method possible given our choice to use
images from the {\em Hubble Space Telescope} or {\em HST} (requiring us to deconvolve the PSF before
shearing and convolving with the HSC PSF).

\subsection{Shear estimation}

The goal of this paper is to test our shear estimation process in
the HSC survey and build a systematic error budget for effects such as multiplicative calibration
bias, which cannot be determined directly from the data.  Here we briefly summarize the shear estimation
software that was used to produce the first-year HSC shear catalog described in
\citet{2018PASJ...70S..25M}.

Galaxy shapes are estimated on the coadded $i$-band images using the {\sc GalSim} implementation of the re-Gaussianization PSF
correction method
\citep{2003MNRAS.343..459H}.  The results of this moments-based PSF correction method are the components of the distortion,
\beq
(e_\mathrm{1},e_\mathrm{2}) = \frac{1-(b/a)^2}{1+(b/a)^2}(\cos 2\phi, \sin 2\phi),
\label{eq:e}
\eeq
where $b/a$ is the axis ratio and $\phi$ is the position angle of the major axis with respect
  to the equatorial coordinate system.  The ensemble average distortion is then an
estimator for the shear $g$, with the process discussed in more detail in Section~\ref{subsec:calib}.

It is useful to be able to apply selection criteria based on
how well-resolved the galaxy is compared to the PSF.  For this purpose, we use the resolution factor $R_2$,
which is defined using the trace of the moment matrix of the PSF
$T_\mathrm{P}$ and of the observed (PSF-convolved) galaxy image
$T_\mathrm{I}$ as
\beq\label{eq:def-r2}
R_2 = 1 - \frac{T_\mathrm{P}}{T_\mathrm{I}}.
\eeq
Well-resolved objects have $R_2\sim 1$ and poorly-resolved
objects have $R_2\sim 0$.

Another useful galaxy quantity is the signal-to-noise ratio of the detection, S/N.  For this
purpose, we use the S/N of the unforced $i$-band \texttt{cmodel}\footnote{`\texttt{cmodel}' refers to composite model photometry that is
      estimated by fitting a linear combination of an exponential profile and de Vaucouleurs profile
      convolved with the PSF model to object light profiles
 \citep{Lupton:2001} as described in \citet{2018PASJ...70S...5B}.} flux.

\section{Simulation inputs}\label{sec:siminputs}

To produce the simulations, we require a set of inputs that specify observing conditions (PSFs,
noise model) and the input galaxy population.  Our basic approach to these was as follows.  For
observing conditions, since all galaxies in a simulated subfield have the same PSF and noise model,
we made each simulated subfield correspond to some random point from the real data.  In the limit of
a very large number of simulated subfields, the distribution of observational conditions in the
simulations will then match the data overall.  For the galaxy population, we use image cutouts from
the {\em HST}, selected as described below.

\subsection{PSFs}

As described in \citet{2018PASJ...70S...5B}, the HSC pipeline separately models the PSF as a function of position in
individual exposures, then builds up a model for the PSF in the linear coadd from the PSF model in
each contributing exposure.  As inputs to the simulations in this work, we used the coadd PSF model
constructed in the pixel basis at random locations within the HSC survey footprint.  A given PSF
model image became the PSF for an entire simulated subfield with $10^4$ galaxies.

\subsection{Noise model}\label{subsec:noise}

For the majority of the galaxies used for shear estimation in HSC, the noise due to the sky level dominates
over the noise due to the object flux, and the sky level is high enough that the Poisson noise is
effectively Gaussian.  As a result, our starting point for the simulations is
to take estimates of the sky variance including all contributions (shot noise, read noise, etc.)
from randomly-distributed locations in the HSC survey.  We selected 200 random positions from each
processed HSC patch (squares with dimensions $\sim$12$\times$12\arcmin) and examined the box of
21$\times$21 pixels around each point.  If at least 95$\%$ of the pixels in a region were not
flagged as being associated with a detected object, we include this region, otherwise it is
rejected.  Fully observed patches typically had $\sim$80 of the 200 points selected.  The variance
per pixel in the coadded HSC images was computed for each blank sky region that was selected. 

\begin{figure}
	\begin{center}
		\includegraphics[width=0.5\textwidth]{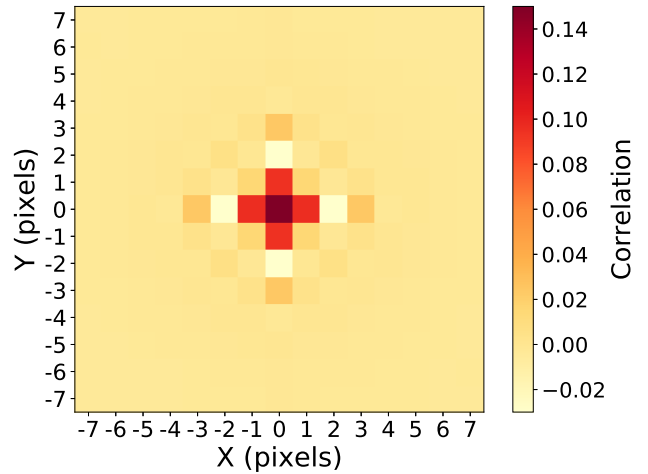}
	\end{center}
	\caption{The noise correlation function as a function of distance, averaged over all the blank
      sky regions.  The color scale is truncated to increase the dynamic range; the correlation is
      defined such that the pixel at $(0,0)$ has a value of 1. \label{fig:noisecorr}}
\end{figure}

However, the resampling process to produce the HSC coadds 
induces pixel-to-pixel correlations which do not vary strongly across the sky.  We use
the cutout blank sky regions to characterize the average noise correlation function, and use the
same noise correlation function for all simulations (see Fig.~\ref{fig:noisecorr}), simply rescaling the overall variance to
match the estimates from randomly-distributed survey locations.  The randomly-distributed survey
locations used for this purpose include some areas that do not pass the weak lensing ``full depth
full color'' (FDFC) cut; we account for this mismatch in a later step of the analysis (Section~\ref{subsec:reweights16a}).

\rev{The dominant factor that determines the correlated noise patterns due to interpolation to a
  common pixel grid is simply the interpolation kernel, which is the same everywhere.  We have
  explicitly confirmed that if we consider different patches in the survey, and find the average
  spatial correlation function of the noise across an entire patch, then each patch gives very
  similar results.  Smaller-scale variation of the noise correlation function is difficult to
  measure empirically, so we estimated its magnitude on theoretical grounds: for a given point in
  the coadd, we find the single visits and their corresponding WCS and noise properties.  After
  coadding many noise realizations of the visit, it is possible to calculate the average noise
  correlation function from the different realizations.  Since the dominant correlations are on
  distance scales of 1 pixel, we focus just on the value of the noise correlation function at that
  separation, finding that it ranges from 0.08--0.11 in the majority of the area of 
  each patch, though with occasional excursions down to 0.06 or up to 0.13.  In
  Section~\ref{subsec:results-shearcalib}, we consider the implications for our shear calibration
  estimates due to the fact that we have neglected this spatial variation of the noise correlation function.}

\rev{In order to avoid propagating the full noise covariance to the coadd, the HSC pipeline scales
  the variance values stored with the coadded image such that the measured variance of the quantity
  (signal/standard deviation) is 1 for pixels identified as noise pixels containing no astrophysical
  object.  In principle the goal is that photometry algorithms would then report approximately the
  same uncertainty that they would measure if the full covariance and true variance were available,
  though this is unlikely to be true in detail.  The scaled variance is also used when measuring shapes and estimating
  the shape measurement uncertainty, and hence it must be included in our simulations so we can
  reverse its effect while accounting more rigorously for the effects of correlated noise.}  The simulation estimates of variances (of fluxes etc.) must be
similarly rescaled using the same rescaling factors as in the real data in order for the
catalog-level outputs in data and simulations to be comparable.

\subsection{Galaxy models}\label{subsec:galaxymodels}

Multiple different parent samples as described below and summarized in Table~\ref{tab:parentsamp} were used to make several sets of simulations.
These different simulation sets were used to explore different issues.
\begin{table*}
\caption{Summary of different parent galaxy samples from Section~\ref{subsec:galaxymodels},
  including how the samples were selected; how the galaxy model is represented in the simulations; whether nearby galaxies are included in the simulations or not; whether the 7\% of
  the objects in the sample that have possible image artifacts, poor masking, or bad parametric fits
  are eliminated; and whether or not the HSC pipeline analysis included detection and deblending.
  See the relevant 
  section listed in the table for more detail.  The majority of our analysis uses parent sample 4,
  with the other 3 only used to understand the sensitivity to various steps in the analysis.}\label{tab:parentsamp}
\begin{center}
\begin{tabular}{lllllll}
\hline
Name & Section & Selection & Galaxy model & Includes nearby & Marginal & HSC pipeline \\
 &  & & & structure? & cuts? & analysis \\ \hline
1 & \ref{subsubsec:deblps} & {\em HST} selection \& cuts \& deblending & Image & No & Yes & No detection, deblending \\ \hline
2 & \ref{subsubsec:paramfits} & {\em HST} selection \& cuts \& deblending & Parametric fit & No & Yes & No detection, deblending  \\ \hline
3 & \ref{subsubsec:nodeblps} & {\em HST} selection \& cuts & Image & Partially & No & Detection, deblending  \\ \hline
4 & \ref{subsubsec:finalps} & HSC selection & Image & Yes & No & Detection, deblending \\ \hline
\end{tabular}
\end{center}
\end{table*}

In all cases, the COSMOS survey is the source of the parent samples.  
The COSMOS {\em HST} Advanced Camera for Surveys (ACS) field
\citep{2007ApJS..172..196K,2007ApJS..172....1S,2007ApJS..172...38S} is
a contiguous 1.64 degrees$^2$ region that was tiled by 575 adjacent and slightly
overlapping pointings of the ACS Wide Field Channel. Images were taken through the
wide F814W filter (`Broad I'). In this paper we use the
`unrotated' images (as opposed to North up) to avoid rotating the
original frame of the PSF. The raw images were corrected for charge transfer
inefficiency (CTI) following \citet{2010MNRAS.401..371M}. Image
registration, geometric distortion, sky subtraction, cosmic ray
rejection and the final combination of the dithered images were
performed by the multidrizzle algorithm \citep{2002hstc.conf..337K}. As
described in \citet{2007ApJS..172..203R}, a finer pixel scale of $0.03\arcsec/$pix was
used for the final co-added images. The 
catalog of object detections was constructed following the methodology
outlined in \citet{2007ApJS..172..219L} and then matched to the COSMOS
photometric redshift catalogue presented in \citet{2009ApJ...690.1236I}.

\subsubsection{Deblended postage stamps}\label{subsubsec:deblps}

The parent sample called `1' in Table~\ref{tab:parentsamp} was constructed in the same way as the
parent sample used for the GREAT3 challenge \citep{2014ApJS..212....5M}, following 
\citet{2012MNRAS.420.1518M}, but with a deeper magnitude limit of F814W$<25.2$.  In short, the object
selection proceeded from a catalog of objects in the COSMOS survey described at the start of this
subsection.  In addition to the magnitude cut, a set of selection criteria were imposed:
\begin{itemize}
\item MU\_CLASS $=1$: This requirement uses the relationship
    between the object magnitude and peak surface brightness to select
  galaxies, and to reject stars and junk objects such as
    residual cosmic rays \citep{2007ApJS..172..219L}.  
\item CLEAN $=1$: This cut eliminates galaxies with defects due
 to very nearby bright stars, or other similar issues.
\item GOOD\_ZPHOT\_SOURCE $=1$: This cut requires that there
  be a good photometric redshift in the COSMOS catalog.  More specifically, it requires that there
  is a match between the detection in the {\em HST} imaging and the Subaru $BVIz$ ground-based catalog, that the object
  not fall within a mask in the ground-based imaging, and that the estimated photometric redshift
  lies in the range $[0,9]$.
\end{itemize}
A random subsample of these galaxies was selected to reduce the size of the dataset.

For each of these galaxies, the size of the postage stamp was estimated based on the object size
as described in \citet{2012MNRAS.420.1518M}.  If this postage stamp size causes the
postage stamp to hit the CCD edge, then the galaxy is eliminated.
Consequently, the probability of a galaxy in our
parent sample having a postage stamp is a weak function of the observed
galaxy size, specifically the FLUX\_RADIUS determined by
SExtractor\footnote{\url{http://www.astromatic.net/software/sextractor}}
\citep{1996A&AS..117..393B}.  Fortunately, it is possible to reweight the sample probabilistically
to counteract this selection effect when making simulations.

Finally, pixels flagged by SExtractor as belonging to some other object were masked with a
correlated noise field.  This means that we rely on the entire object detection, selection, and
deblending process in COSMOS when creating the sample.  The last of these is questionable in the
context of simulating a ground-based survey, because it may be possible to deblend two objects in
space-based imaging that would always be detected as just a single object from the ground.  Thus, in
the simulations, the system appears as two distinct objects, whereas in the HSC data it would always
appear as just one due to unrecognized blending.  In this case, if the two objects are at the same
redshift, then the right thing to do when making simulations to test shear recovery is to simply
include it as a single system.  If they are at different redshifts, then their true shears differ
and it is unclear how to interpret the inferred calibration biases from simulations that assume they
are at the same redshift.

This parent galaxy sample is publicly
available\footnote{\url{http://great3.jb.man.ac.uk/leaderboard/data/public/COSMOS\_25.2\_training\_sample.tar.gz}}
and is set up for easy use with \galsim.  The \galsim\
\texttt{COSMOSCatalog} class handles the needed numerical manipulations such as removal of the {\em HST}
PSF, shearing in Fourier space, convolving with the target PSF, and resampling to the target pixel
scale in a way that is transparent to the user.  This process was numerically validated to the level
needed to test shear at the level needed for Stage IV dark energy surveys, as described in
\citet{2015A&C....10..121R}.

 The \texttt{COSMOSCatalog} class also optionally imposes cuts on the catalog to
eliminate objects with possibly problematic image postage stamps (e.g., contaminated by image
artifacts) or parametric fits using a keyword \texttt{exclusion\_level=`marginal'}.  This cut 
eliminates $\sim$7\% of the objects, and was used for sample 1.  Many of the problematic
postage stamps have issues with masking of nearby objects, which is the primary motivation for using
this cut even though the sample 1 simulations use the image rather than a parametric model.

However, it was realized in the process of carrying out the work in this paper that imposing the GOOD\_ZPHOT\_SOURCE
$= 1$ selection criterion had the unintended consequence of excluding galaxies with close
neighbors. We  find  that this flag induces a separation-dependent pair exclusion for galaxy pairs
closer than 2.4\arcsec, 
down to 0.9\arcsec where all nearby galaxy pairs \rev{in which both galaxies are above the detection
limit} are effectively removed by this  cut. 
There are several possible reasons for this problem: poor deblending in the ground-based
catalog, photometric redshift failures preferentially for close pairs (though this explanation is
disfavored by the findings of \citealt{2015MNRAS.449.2128K}), or over-deblending in the {\em HST} image
processing. The potential exclusion of close galaxy 
pairs in the parent sample used for our simulations may be problematic for the purpose of this work,
leading to an under-representation of the population of blended galaxies \rev{for which both objects
are above the detection limit}. \rev{For objects that are too close to be distinguished, the fact
that both are not in our catalogue is actually a positive side effect from this selection criterion;
we only want that unrecognized blend system consisting of the two galaxies together to appear in simulations once, not twice (which would lead to
double-counting).  However, while there are some advantages and disadvantages to this selection
criterion, the desire for more control over how galaxies are selected} led us to redefine a parent sample defined based on actual HSC detections in the COSMOS field (parent sample 4, described below).

\subsubsection{Parametric fits}\label{subsubsec:paramfits}

Parent sample 2 is essentially the same as sample 1, except that instead of using the {\em HST} images as
the basis for the image simulations, we use a set of \sersic\ fits to those {\em HST} images.  The fitting
methodology is the same as was used for the GREAT3 challenge (described in detail in
\citealt{2014ApJS..212....5M}).  In short, galaxies were fit to a single \sersic\ profile and to a
sum of a bulge and disk component with fixed \sersic\ indices.  For those galaxies for which a
two-component fit was statistically preferable \rev{as determined by its having a lower median
  absolute deviation (or MAD)}, that fit was used, while for the others ($\sim 70$\%
of the sample) the single \sersic\ fit was used.  The \galsim\ \texttt{COSMOSCatalog} class is set
up to interchangeably use the parametric fits that are publicly distributed with this sample (same
URL as for sample 1); a single keyword argument determines whether the {\em HST} image or the parametric
fit is used.  Because of this, the only difference between simulations with sample 1 and sample 2 is
the inclusion of parametric (sample 2) or fully realistic (sample 1) galaxy morphology.


\subsubsection{Postage stamps without deblending}\label{subsubsec:nodeblps}

Parent sample 3 is very similar to parent sample 1, but in this case, the nearby objects detected
based on the SExtractor deblender were {\em not} masked with a correlated noise field.  Hence sample
3 includes not only realistic galaxy morphology, but also the impact of nearby structure and
unrecognized blends.  In Table~\ref{tab:parentsamp}, this sample is listed as `partially' including
the effects of nearby structure.  The reason this is partial rather than complete is that some of
the postage stamps used were still relatively small -- small enough that they do not cover the
entire postage stamp area in the simulations.  In principle this is unlikely to be a problem,
because the outer edges contain objects that would always be securely detected and deblended in the
ground-based survey, but we cannot {\em a priori} rule out low-level issues due to these small-area postage stamps.

Because many of the image artifacts that are flagged with the `marginal' cut relate to masking of
nearby objects, and those objects are deliberately not masked for sample 3, the `marginal' cut was
not used for this sample.

\rev{Note that the comments about nearby galaxy pairs at the end of Section~\ref{subsubsec:deblps} for parent sample
1 also apply here, since parent sample 3 was selected in the same way as parent sample 1.}

\subsubsection{Final parent sample}\label{subsubsec:finalps}

Parent sample 4 is the one that we use for the majority of the results in this paper.  Its design
was intended to serve several purposes that improve upon the pre-existing samples 1--3:
\begin{itemize}
\item To avoid selection effects due to the selection criteria for samples 1--3 described in
  Section~\ref{subsubsec:deblps}, we select the galaxies based on HSC detections with minimal
  constraints, rather than selecting them based on {\em HST} detections with additional constraining
  selection criteria.  The process of selecting based on HSC will be described below.
\item We do not deblend based on the {\em HST} images.  Instead, the images are fed into the simulation
  software and used to make simulations, and we rely exclusively on the HSC deblender (as 
  for real analysis of HSC data, and simulations with parent sample 3).
\item We use the real {\em HST} galaxy images (as in samples 1 and 3), not parametric fits (as in sample 2),
  to include realistic galaxy morphologies.
\item The postage stamp sizes are chosen such that all structure around the central detection out to
  the edge of the simulated postage stamps is included in the simulations.
\end{itemize}

\rev{This sample has been publicly distributed as part of an HSC survey incremental data release\footnote{\url{https://hsc-release.mtk.nao.ac.jp/doc/index.php/weak-lensing-simulation-catalog-pdr1/}}.}

Our selection of parent sample objects begins with three HSC Wide-depth coadds in the COSMOS region
that were described in \citet{2018PASJ...70S...8A}.  These coadds were produced from the HSC Deep
layer data, and have fewer exposures than in a typical Wide layer coadd due to the differences in
exposure times.  Exposures with different seeing values were used to create Wide-depth stacks with
effective seeing better than, around the median of, and worse than the HSC Wide layer on
average\footnote{After this work was completed, a problem with the median-seeing coaddition was
  identified, as described on
  \url{https://hsc-release.mtk.nao.ac.jp/doc/index.php/known-problems-in-dr1}, which resulted in
  that coadd being shallower than the others and than the actual Wide-layer coadds by about 0.16
  magnitudes.  Since our $i<24.5$ cut is quite far from the detection limit, 
we do not anticipate that this issue with the coadd causes a serious problem for the
  results presented in this paper.}.  These stacks have seeing values of 0.5\arcsec, 0.7\arcsec, and
1.0\arcsec, as described in Section~3.8 of \citet{2018PASJ...70S...8A}.

For each of the three coadds, we apply the conservative bright object mask described in
\citet{2018PASJ...70S..25M} (not the later version described in \citealt{2018PASJ...70S...7C}).  We
then identify all HSC galaxy detections satisfying the following conditions: basic flag cuts for
object detection, measurement, and star/galaxy classification as given in the top section of Table 4
in \citet{2018PASJ...70S..25M}, followed by a loose cut on magnitude at $i<25.2$ in the HSC imaging. There is no
way to impose the weak lensing full depth and full color (FDFC) cut described in
\citet{2018PASJ...70S..25M}, because of the way these coadds were produced.  The
justification for this magnitude cut is that scatter in the magnitudes between detections of the
same object in the different coadds is large enough that at $2\sigma$, some objects brighter than
$i=24.5$ (our limit used for the shear catalog in the data) might actually come from COSMOS objects
with F814W$\approx 25.2$. We do not want to apply a stricter magnitude cut, because then we would
miss some of the faint objects that may be scattering into our sample in reality. 
We applied no resolution or S/N cut because we found that there was no
reasonable cut that would avoid cutting into the sample of objects in our shear catalog. 
Fortunately, our magnitude cut is sufficient to remove a large amount of junk detections at very low
S/N that would never be used for science, so that we are not resimulating large samples of useless
objects.  

After imposing these cuts, the locations of these HSC detections are compared with the masks on the {\em HST}
COSMOS images, and any locations that are masked or are outside the {\em HST} COSMOS imaging area get
removed.  The locations of the remaining detections will be used as the center of the parent sample
postage stamps, regardless of whether or not there is a COSMOS detection near that point.

The sizes of the postage stamps are determined based on the following criteria: the minimum postage
stamp size is such that the postage stamps cover the entire area of the postage stamps in the
simulations, i.e., $64\times 0.168$\arcsec\ in length.  Given the 0.03\arcsec\ pixel scale for the
resampled COSMOS images, our minimum postage stamp size is $380\times 380$.  For each galaxy, we use
the following formula to determine the nominal postage stamp size needed to contain the galaxy
light: ideal stamp width is $12\times r_\text{det}^{\text{(HSC)}}$, where the radius here is the
observed (PSF-convolved) determinant radius of the HSC detection.  If that ideal width is less than
380, we use 380.  If the ideal width exceeds 1000, we discard the object from our parent sample.
Finally, if the calculated postage stamp size causes the image to overlap a CCD edge, we also
discard it.  This final cut results in a slightly non-representative distribution of galaxy sizes in
the parent sample, with the probability of inclusion in the sample varying by approximately 12 per
cent across the full range of galaxy sizes in the sample.  However, when randomly drawing from the
parent sample, we probabilistically 
reweight the sample to cancel out this selection effect.

Because of our arbitrary choice of postage stamp sizes, it is entirely possible that some
neighboring objects in the postage stamp (not the central one) 
will be cut off by the postage stamp edges.  Fortunately these objects are reasonably far (at least
$5$\arcsec) from the center of the simulated postage stamps, so in general those detections are
being discarded anyway.

While we did not require an {\em HST} COSMOS detection matching the locations of these postage stamps, we
do cross-match against the COSMOS catalog to identify the nearest COSMOS detection.  The purpose of
this matching process is to define {\em HST} PSF models for all the galaxies, so that \galsim\
can remove it as the first part of the simulation process.

The outcome of the above process is a set of three parent samples for the three seeing ranges, assembled
completely independently, though in practice they have a large fraction of objects in common.  
In the simulation scripts, we use a different parent sample for each
subfield depending on the seeing value in that subfield.  For PSF model FWHM values in the ranges
$<0.6$\arcsec, $[0.6\arcsec,0.85\arcsec]$, $>0.85$\arcsec, we use the parent samples defined using
the coadds that have seeing better than, around the median of, and worse than the HSC Wide layer.
Given the seeing distribution for the galaxies in our shape catalog \citep{2018PASJ...70S..25M}, in
practice we almost exclusively use the first two parent samples.

\begin{figure}
	\begin{center}
\includegraphics[width=0.5\textwidth]{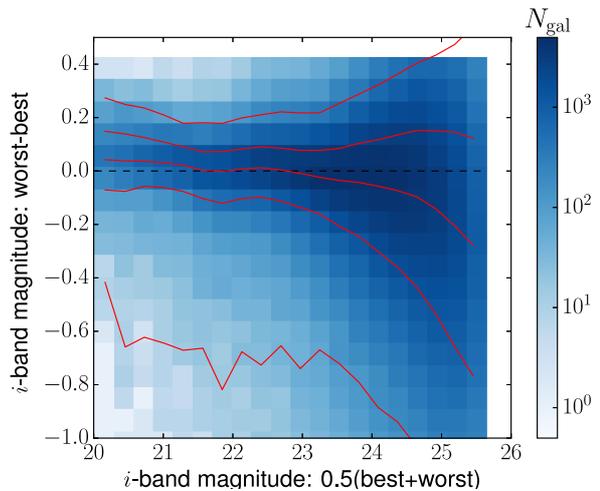}
	\end{center}
	\caption{The logarithmic color-scale shows a 2D histogram of $\Delta$mag (the difference in
      observed magnitude in the worst-seeing coadds vs.\ the best-seeing coadds for objects detected
      in both) as a function of the \rev{average of the observed magnitudes in the worst- and best-seeing coadds}. The five red
      lines show the $(5, 16, 50, 84, 95)$ percentile values of $\Delta$mag, while the dashed black line shows the
      ideal value of 0.  The coherent skew towards negative values indicates that a subset of the
      galaxies in the worst-seeing coadd appear consistently brighter than the same galaxies in the
      best-seeing coadd, due to unrecognized blending effects.\label{fig:compcoadds}}
\end{figure}

The three parent samples contain 216426, 199461, and 184217 objects as we go from best to worst
seeing.  There is a significant overlap in the RA/dec positions, but we do see seeing-dependent
blending effects. An illustration is shown in Fig.~\ref{fig:compcoadds}.  Here we identify
matches within 1\arcsec\ in the best-seeing and worst-seeing coadds, then show a 2D histogram of the difference in
magnitudes for those objects as a function of the magnitude in the best-seeing coadds.  As shown,
there is substantial skew towards negative $\Delta$mag, meaning that the objects
appear brighter in the worse-seeing coadds than in the best-seeing coadds.  This could result from
unrecognized blends in the worse-seeing coadds that are properly recognized and deblended in the
best-seeing coadds.

While we believe the procedure in this subsubsection is a more principled way to create the parent
samples for these simulations, ultimately the justification of this procedure rests on what we find
when analyzing the simulations. That is, the distributions of simulated galaxy properties (as
measured by the HSC pipeline after running the object detection, deblending, and measurement
routines) should match those in the real shape catalog.  In addition, we have no process for
addressing systematic errors due to unrecognized blends at different redshifts, since our simulations
assume they are at the same redshift.

\subsubsection{Impact of different parent sample selection}

\begin{figure*}
	\begin{center}
		\includegraphics[width=3.2in]{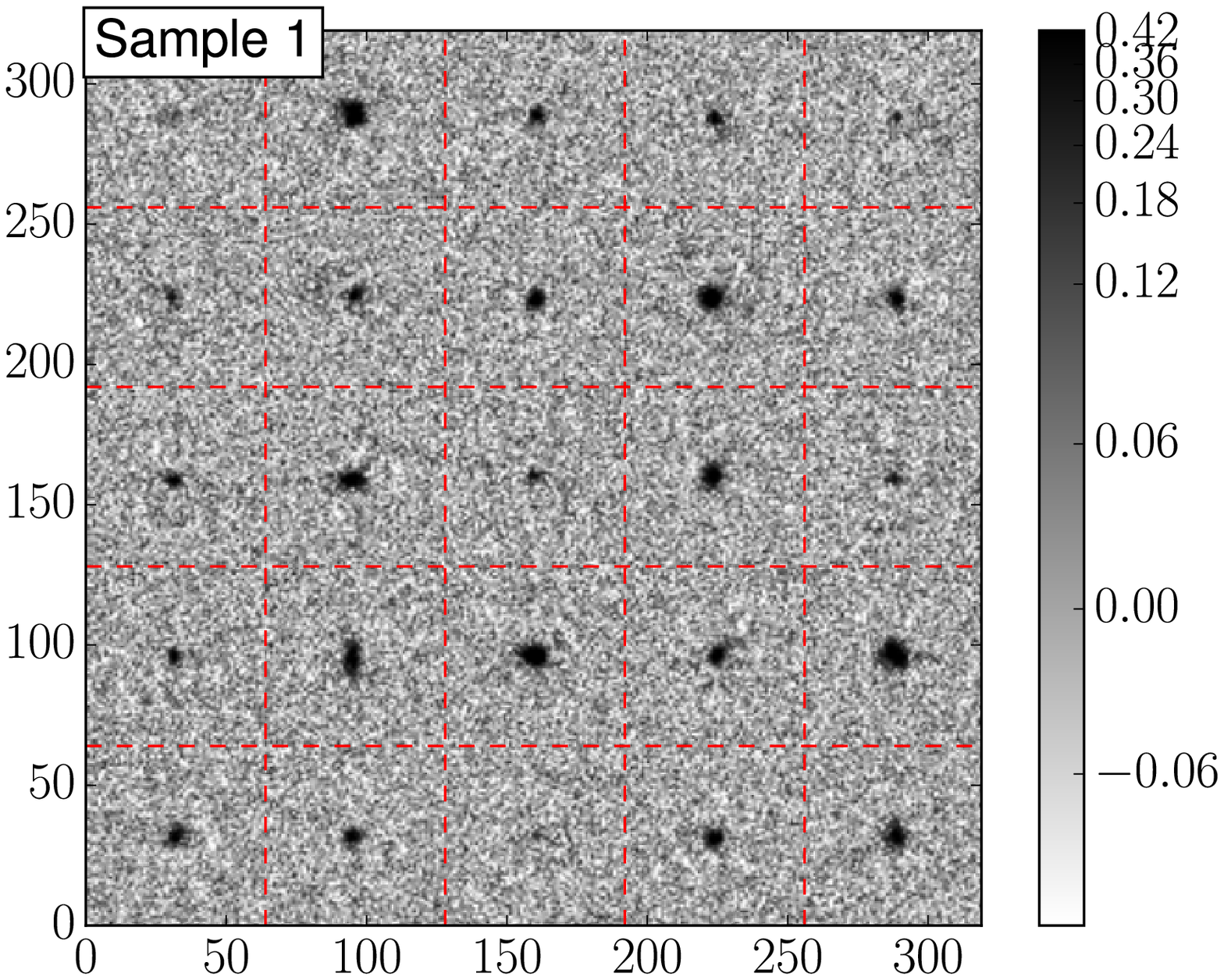}
		\includegraphics[width=3.2in]{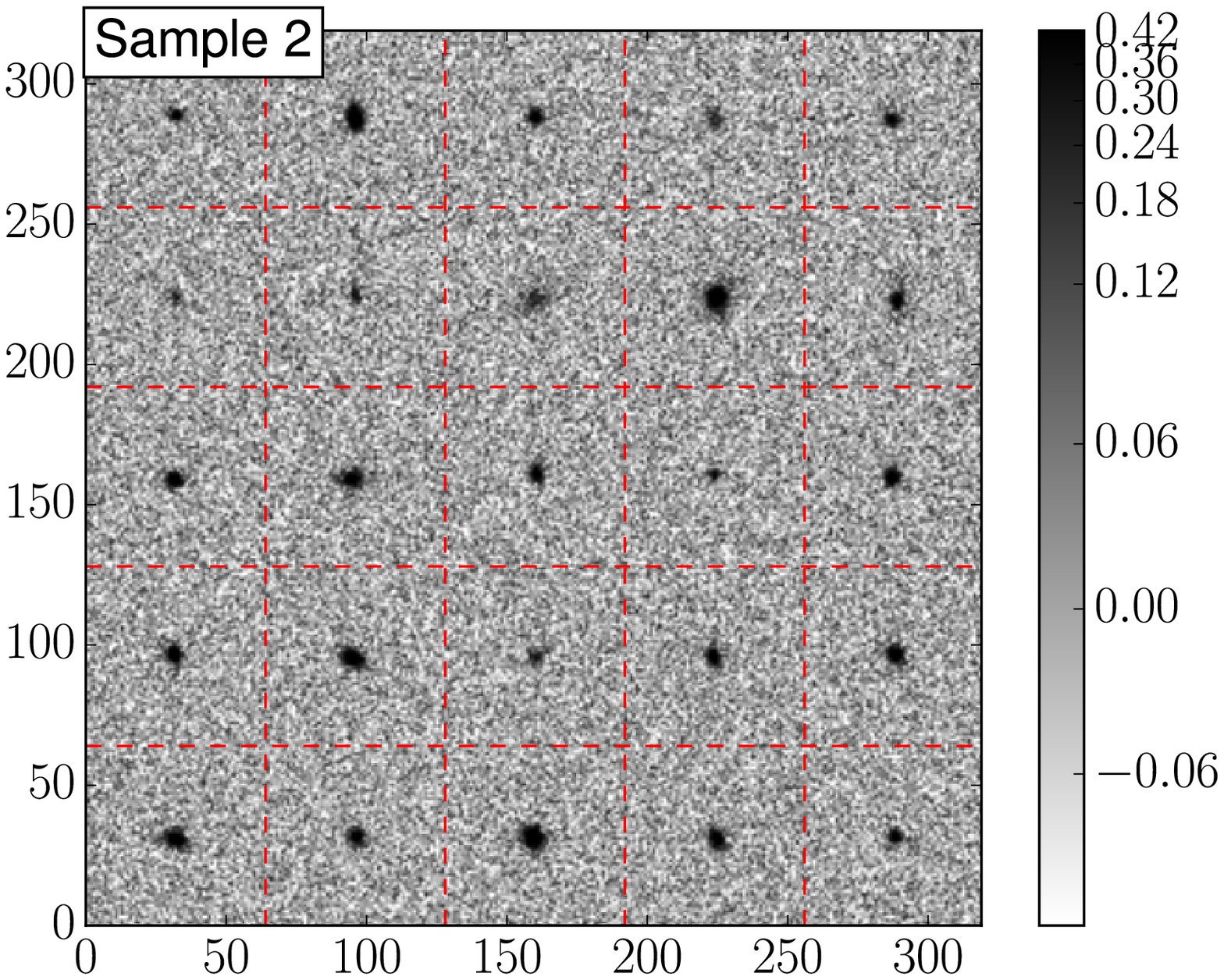}
		\includegraphics[width=3.2in]{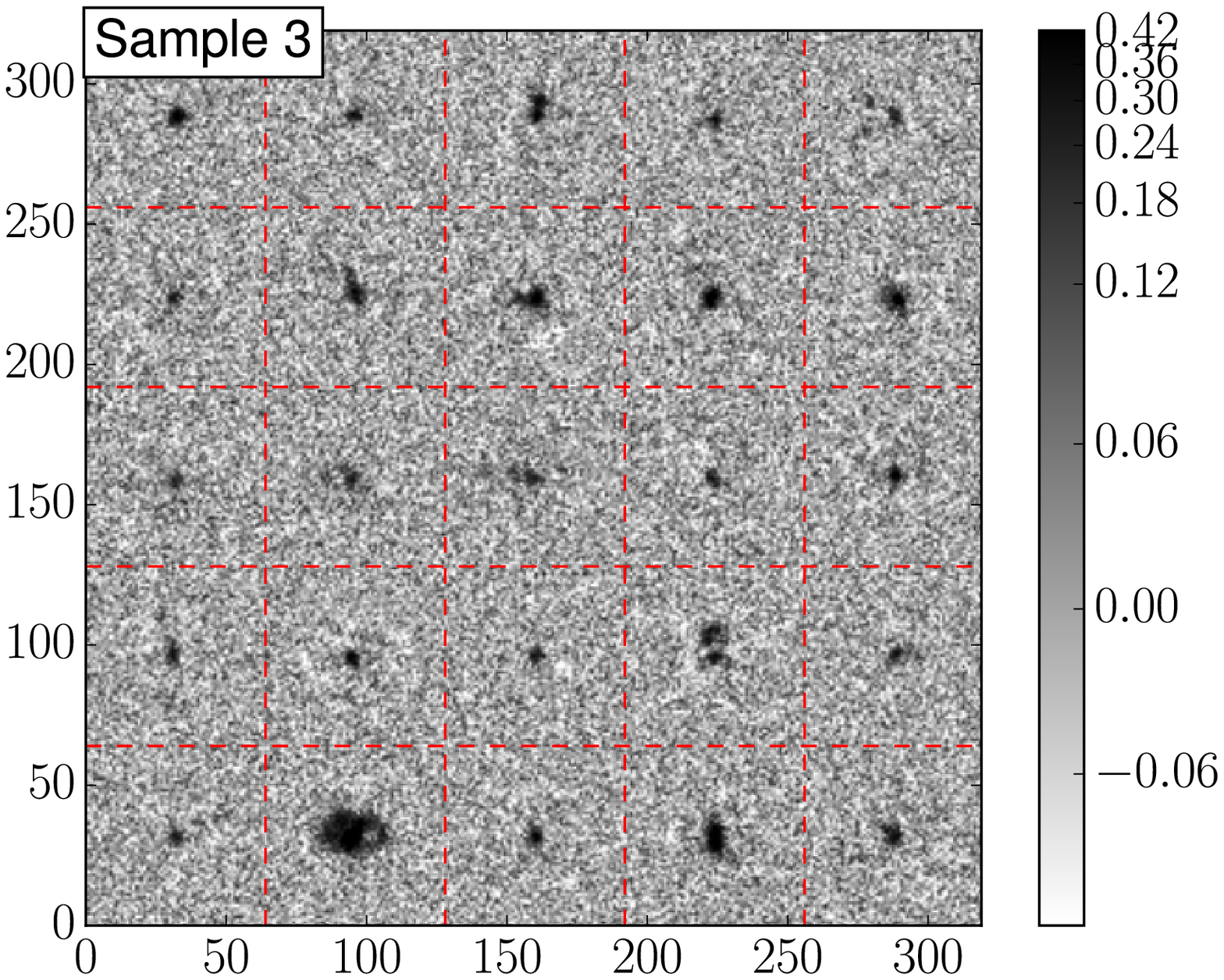}
		\includegraphics[width=3.2in]{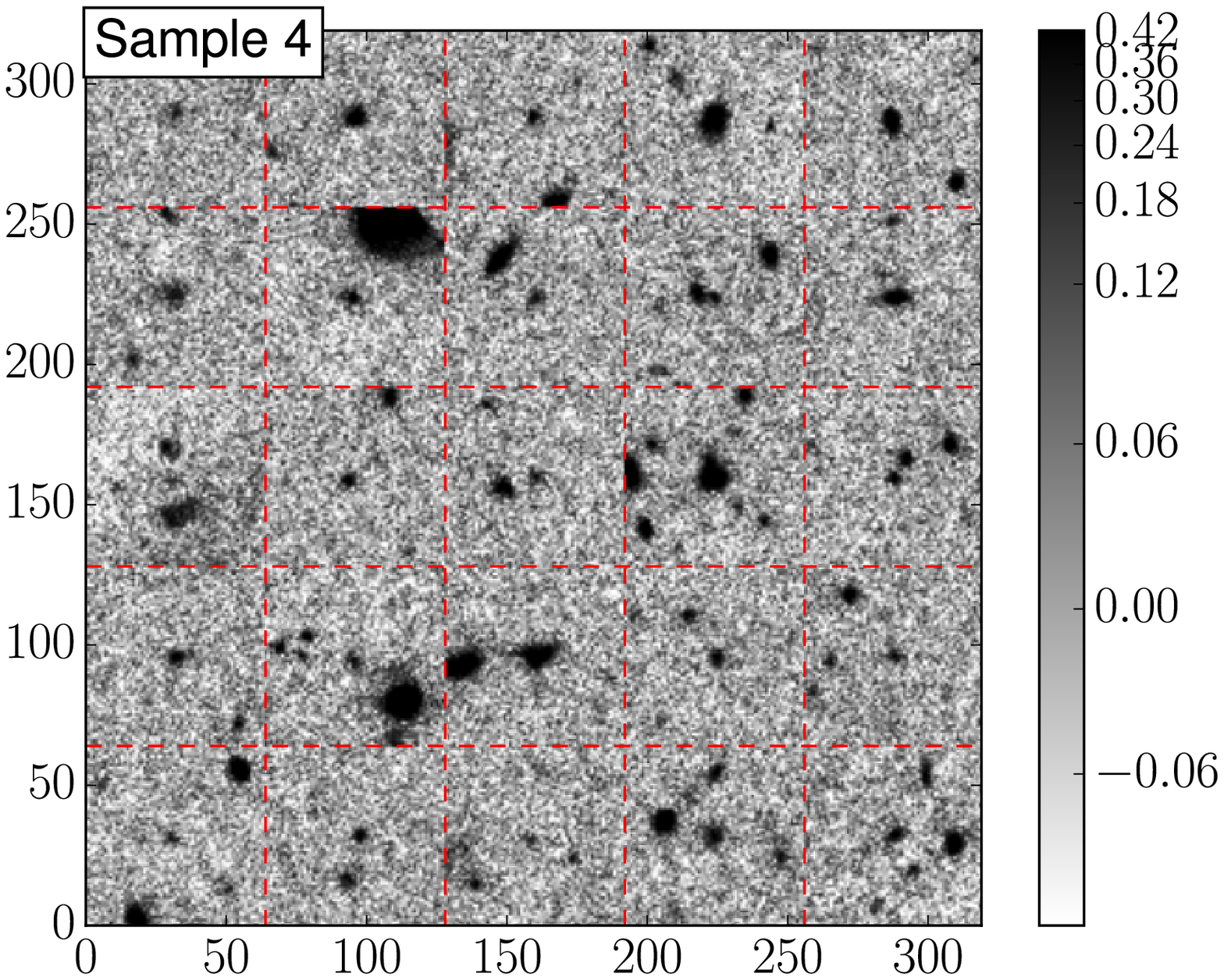}
	\end{center}
	\caption{Example simulated images for simulations with the four parent samples as labeled on
      each plot.  All images have the same zero points, and the symlog color scales are the same on each panel.  For the purpose of
      illustration we have chosen images with the same PSF model.  Dashed red lines show the
      artificial boundaries between individual postage stamps; we have shown a $5\times 5$ postage
      stamp region ($320\times 320$ pixels) out of the simulated image composed of $100\times 100$ postage stamps.   Note that the color-scale is saturated in some places for sample 4
      simulations, because some of the stamps include bright stars that happen to lie close enough
      to our central objects that they are included in the simulations. The sample 3 and sample 4
      images look quite different from each other, despite the choice to not mask nearby objects in either sample,
      because sample 4 images were created from larger postage stamps (sufficiently large that they
      include some irrelevant structures that would never be blended with the central objects).  We
      demonstrate later in this work that the measured properties of the central
      galaxies in the postage stamps and their shear calibration biases are very similar in samples
      3 and 4 (Fig.~\ref{fig:parentsamp_dist}).\label{fig:simimages}}
\end{figure*}

In Section~\ref{subsec:results-samples} we will quantify the impact of
using these different parent samples on estimated shear biases.  For now, we give a visual demonstration of the impact of
different parent samples by showing cutouts from the simulations with different parent samples in
Figure~\ref{fig:simimages}.  As shown, for parent samples 1-2,
      the simulated postage stamps just have a single object by design. For sample 3, there are
      nearby objects, but nothing near the edges of postage stamps, because the typical postage
      stamp size in COSMOS covered a smaller spatial extent than these postage stamps.  However,
      the nearby structure should be the most important at affecting the measured properties of the
      central object.  Finally, by design the sample 4 images have nearby structures included out
      to the edge of the postage stamps, in some cases artificially truncated by the postage stamp
      edges.  In all cases it is only the central object in the postage stamp that is used for the
      simulation analysis.

A final difference between the simulations with the different parent samples is that samples 1--3
were produced with a S/N$<80$ cut, to focus on the low-S/N regime where shear calibration varies
strongly, whereas sample 4 simulations do not have this cut.  This cut could not be applied in
practice because sometimes faint objects had nearby bright ones, and the bright ones contributed to
the S/N estimate that was computed on the fly while producing the simulations (without deblending), 
so the S/N cut would have led to the exclusion of galaxies well below the cut
value.  Because of the S/N cut being present in some simulation sets and not others, the
figures that characterize and compare galaxy populations have a S/N$<80$ cut imposed on all samples 
(to enable fair
comparison across simulation samples), but the shear calibration results that use sample 4 go above
S/N$=80$.

\section{Analysis}\label{sec:analysis}

In this section we describe how the simulations were analyzed.  They were run through the HSC
pipeline followed by a series of analysis routines aimed at measuring shear calibration
biases.

\subsection{HSC pipeline}\label{subsec:hsc-pipeline}

The HSC pipeline, as described in \cite{2018PASJ...70S...5B}, is being developed for the Large
Synoptic Survey Telescope (LSST; \citealt{2010SPIE.7740E..15A,2015arXiv151207914J}).
Because of the setup of the simulations, we did not need to run the full HSC pipeline. 
  We assumed perfect knowledge of the PSF, background, and
astrometric and photometric calibration.  The first step in processing the simulations was to define
a list of objects and their associated pixels.  The list of all the pixels defined for a particular
object is called a footprint.  When processing simulations made with samples 1 and 2, no explicit
detection was performed because there was only one object per postage stamp.  Instead, we used the stamp
center as the peak and assigned all pixels from the stamp to the footprint.  This resulted in larger
footprints for these samples than in the real data, but had little impact on the actual
measurements.  The noise in these images was found by taking the variance of pixels around the
borders of the postage stamps.


For simulations produced using samples 3 and 4, more than one object was present in most postage
stamps.  We therefore ran the detection algorithm to identify sources.  This algorithm performs a
maximum likelihood analysis to detect all pixels with a 5$\sigma$ threshold, using a Gaussian
smoothing filter matched to the size of the PSF.  Connected pixel regions above threshold are
identified as the footprint and sources are defined as peaks within each footprint.   If more than
one peak per footprint was found, we ran the deblending algorithm, which apportions the flux to the
different objects\footnote{The pixels within the footprint are fractionally assigned to the two
  peaks by the deblender, but all of the pixels outside of the footprint that are not part of a
  different footprint are used when measuring both peaks. As a result, there is some low-level
  double-counting in the measurement, but the Gaussian weighted moments should be downweighting
  those pixels by a large factor.  Any biases that result from this double-counting in real data
  should be reflected in the simulation analysis as well.}.  In the event that no object was detected in a postage stamp, we artificially
inserted one with the peak at the center of stamp.  Because objects could be near an edge in these
samples, the noise for these was calculated by taking the variance of all pixels not part of a
detected footprint.  The final step was to run the full suite of measurement algorithms on the
deblended images.  Table~\ref{tab:parentsamp} includes a summary of the key difference between how
the simulations with the different parent samples were analyzed in the HSC pipeline (i.e., was
detection and deblending done or not). 

The main differences between the coadded $i$-band data and simulations are: 
\begin{itemize}
	\item The presence of artifacts like cosmic rays in the data.
	\item In the data, images are resampled and then coadded.   In the simulations, we generate a
      single-band image and account for the impact of resampling on average by adding correlations to the noise as described in Section~\ref{subsec:noise}.
	\item Object detection is done on the data by merging peaks from multiple bands.  This can lead
      to more spurious objects when making cuts on $i$-band quantities; hence, in the data we have a
      multi-band detection requirement, which is irrelevant in the single-band simulations.
	\item The number of objects in a blend is typically smaller in the simulations than in the data. This difference
     is potentially relevant because the performance of the deblender has been shown to be worse as
     more objects are included.  However, the number of many-object blends in the data is enhanced
     by the presence of bright stars and galaxies; in some cases, none of those will be deblended
     into systems that have any lensing-selected galaxies.  As a result, this difference may not
     result in the lensing-selected samples in simulations and data having any important differences.
	\item The simulations only contain galaxies at the centers of postage stamps. For samples
      3 and 4, images of stars could appear on galaxy outskirts, not at the centers of postage
      stamps; for samples 1 and 2, no stars will appear anywhere.
	\item The data contains relative astrometric offsets between visits slightly blurring the PSF
      relative to the model.  As shown in 
      \citet{2018PASJ...70S..25M}, the impact on the coadded PSFs is negligible, which justifies our
      ignoring this effect in this work.
\end{itemize}
These differences require us to impose slightly different cuts in order to get consistent samples
between the data and simulations, as described below.

\subsection{Higher level analysis software framework}

In order to ensure consistency in the treatment of simulation outputs, the results presented in the 
remainder of this work rely on a common analysis framework. This framework is tasked in particular with
ensuring that simulations match the observed data in terms of noise variance and PSF size, through a 
reweighting scheme described in Section~\ref{subsec:reweights16a}. It also provides a unified set of routines
to transparently handle simulations and data, and apply various cuts and selection criteria in a consistent
fashion.

The simulations are specifically designed to include $90^\circ$ rotated pairs of galaxies
which can be used to nearly cancel out shape noise. This feature is
particularly useful for our estimation of shape measurement errors, as described below, and to reduce
statistical errors on shear biases, enabling them to be robustly quantified with fewer simulations. 
By keeping track of the members of each pairs, the analysis
framework provides several options to apply this cancellation or not. In all cases, a basic set of
flag cuts are imposed.  These are a subset of the cuts in the `Basic flag cuts' section of table
4 of \citet{2018PASJ...70S..25M} for 
the shear catalog, where the cuts that are omitted are unnecessary in the simulations due to how
they were produced.  Specifically, for simulations (which are single-band images with no image artifacts),
we only need the cuts on
\texttt{idetect\_is\_primary}, \texttt{iflags\_badcentroid}, \texttt{icentroid\_sdss\_flags},
\texttt{iclassification\_extendedness}, and \texttt{ishape\_hsm\_regauss\_sigma}.  We also impose a
cut specific to the simulations, requiring that the detection nearest the center of the postage
stamp have a centroid that is a maximum of 5~pixels from the center; this cut was empirically
determined as a way of eliminating stamps where the detection nearest the center was not the
intended central object.

After the imposition of those flag cuts, further selection criteria in the `Cuts on object
properties' section of table 4 of \citet{2018PASJ...70S..25M} are imposed based on whether
shape noise cancellation is desired or not.  Without shape noise cancellation,
all galaxies passing a given set of cuts are retained. With shape noise cancellation, pairs where
one member has not been detected by the HSC pipeline are first 
discarded,  and of the
remaining pairs, only those in which a randomly selected member passes the specified set of cuts are retained,
as to avoid selection bias entirely. 

\subsubsection{Different simulation and data versions}\label{subsec:reweights16a}

\begin{figure*}
\begin{center}
\includegraphics[width=5in]{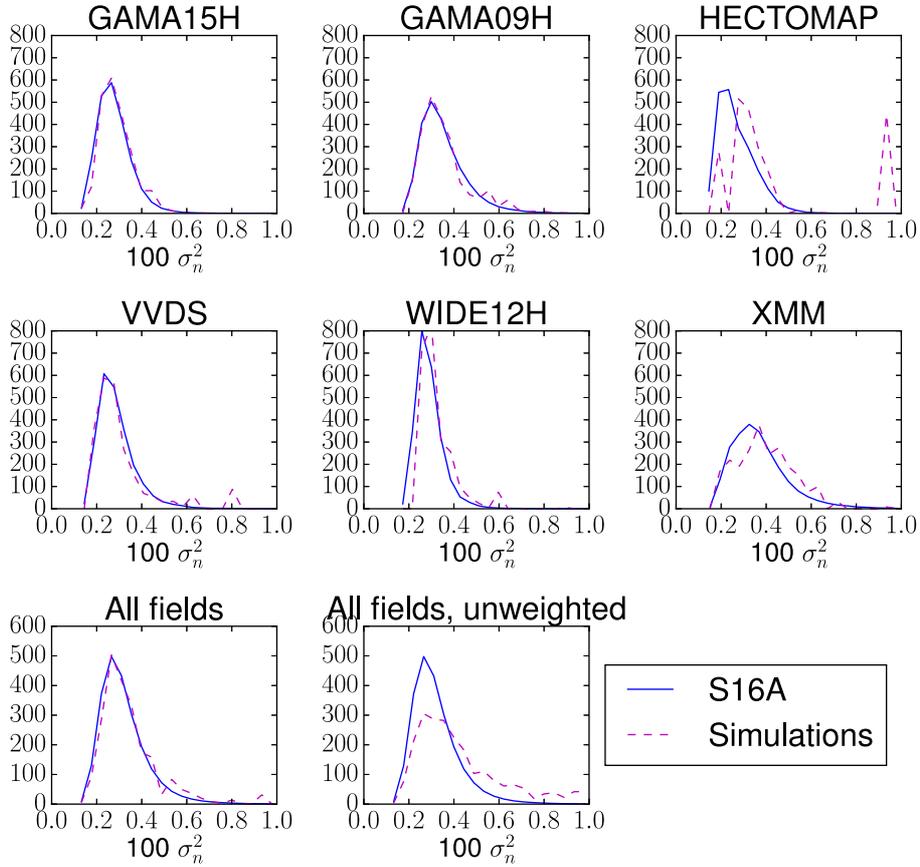}
\end{center}
\caption{Distribution of the noise variance, for each field and for simulations vs.\ data
  (S16A), after reweighting the simulations to match the imaging conditions in the data.  The top two rows are for each field individually, while the bottom left is for the
  average over all fields; all panels show reasonable consistency between simulations and data.  Finally, the bottom middle panel shows the average over all fields, {\em
    without} reweighting to account for differences in imaging conditions.   \label{fig:noise-var}}
\end{figure*}

The simulations described in this work were produced using PSF and sky variance levels corresponding
to an earlier version of the shape catalog compared to the one used for science and characterized in
\citet{2018PASJ...70S..25M}.  The primary differences are that the catalog from
\citet{2018PASJ...70S..25M} includes data that were taken later, but has regions that do not
correspond to full depth in all five bands removed.  As a result, the simulations have some
subfields with much higher noise variance than in the final shape catalog (Fig.~\ref{fig:noise-var}).

To address this issue, we included in our software framework an option to separately reweight the
simulations to match the distributions of sky variance and PSF FWHM as in the final shape catalog.
This results in a small fraction of the subfields being excluded altogether, while many others have
weights that differ from 1.  The reweighting does not account for differences in PSF shape or higher
order moments; however, these should have a relatively smaller effect on the observed galaxy
properties in the simulations and data compared to the sky noise level and PSF FWHM.

\begin{figure*}
\begin{center}
\includegraphics[width=5in]{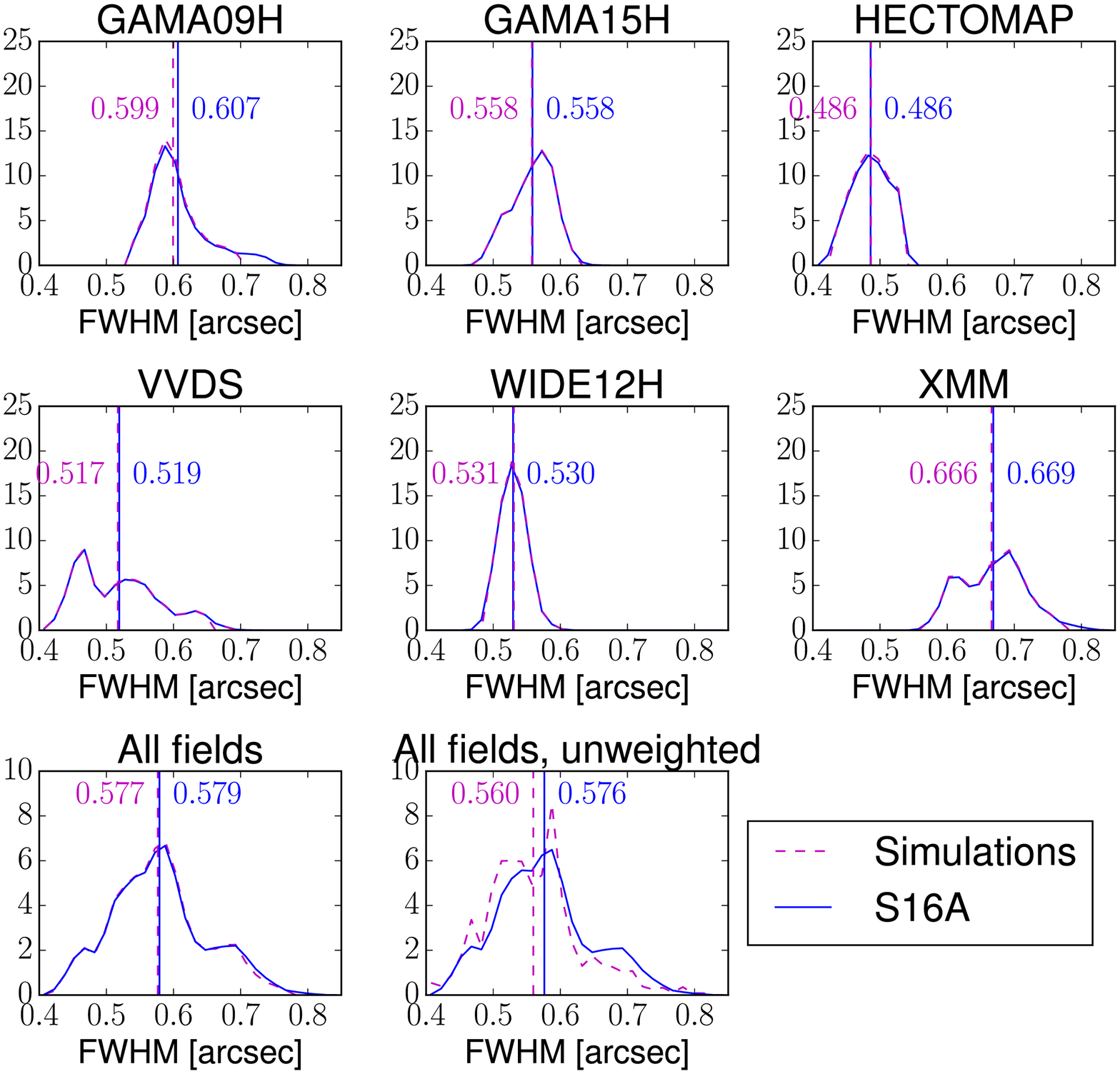}
\end{center}
\caption{Distribution of the PSF FWHM for each field in HSC
  and for simulations vs.\ data  (S16A), after reweighting the simulations to match the imaging conditions in the data.  The top two rows are for each field individually, while the bottom left is for the
  average over all fields; all panels show reasonable consistency between simulations and data.  Finally, the bottom middle panel shows the average over all fields, {\em
    without} reweighting to account for differences in imaging conditions. The weighted mean PSF
  FWHM values are given on the plot and indicated with
  vertical lines. \label{fig:psf-size}
}
\end{figure*}
\begin{figure*}
\begin{center}
\includegraphics[width=5in]{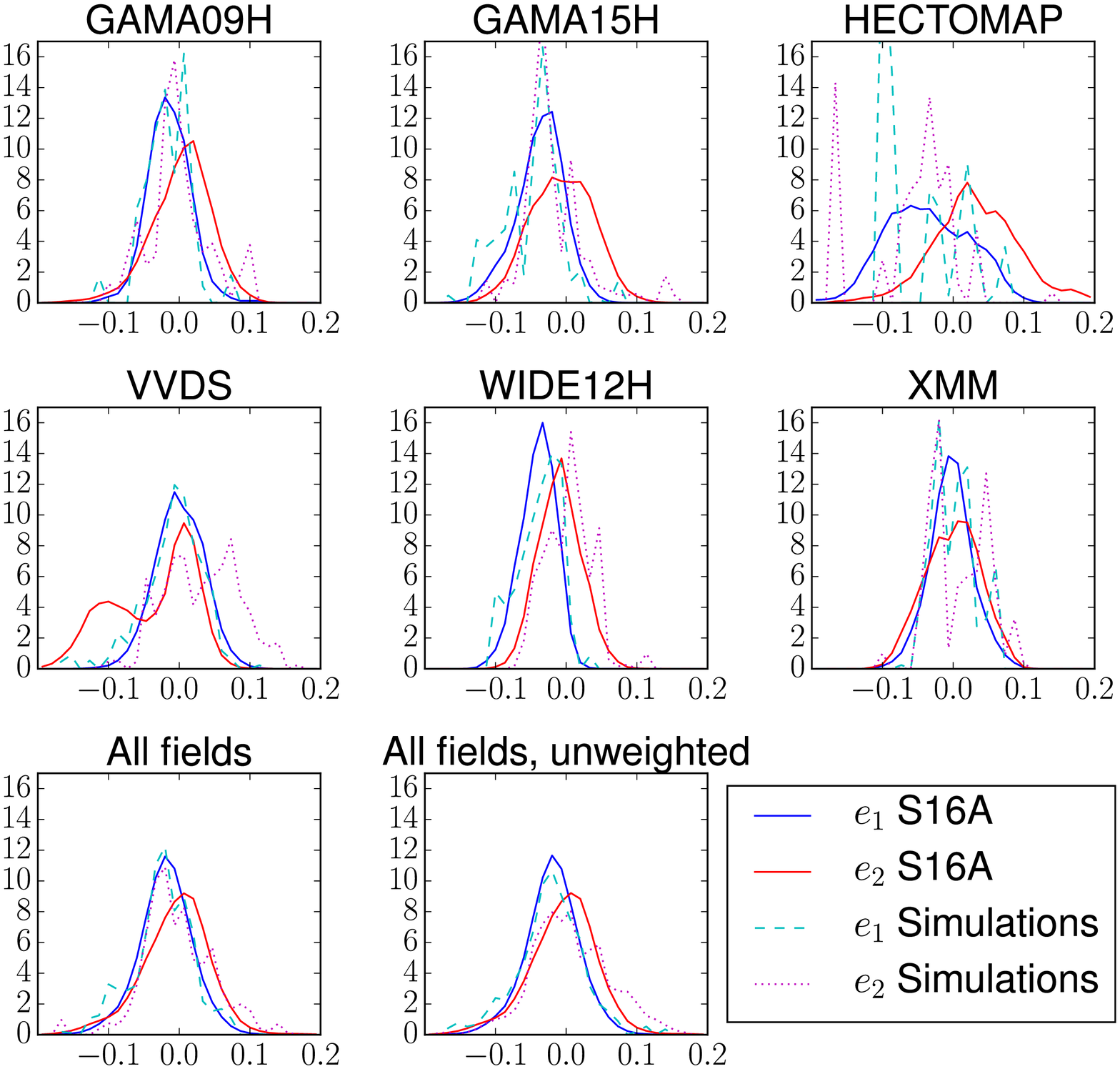}
\end{center}
\caption{Distribution of the per-component PSF ellipticity, for each field in HSC and
  for simulations vs.\ data. The panels are defined in the same way as in Fig.~\ref{fig:psf-size}. \label{fig:psf-e}}
\end{figure*}

As shown in Figure~\ref{fig:psf-size}, after reweighting, the distributions of PSF FWHM in
simulations and data are well-matched.  The match
will not be perfect because of the fact that our simulations have the same PSF within each subfield,
and hence the finite number of subfields is a limitation in matching the full joint distribution of
PSF FWHM and noise variance (Fig.~\ref{fig:noise-var}).  Figure~\ref{fig:psf-size} shows that the
means and overall shape of the PSF size distributions in the simulations and data are indeed
well-matched; for the survey overall, the mean PSF sizes in the data and simulations agree to
within 0.3 per cent, an order of magnitude better than without the reweighting.

We have not made any attempt to reweight the simulations to match the distribution of PSF
ellipticity in the final shear catalog, and hence some mismatch is seen as in
Figure~\ref{fig:psf-e}.  However, the primary reason to do the reweighting is to match the empirical
distribution of object properties that determine shear calibration (such as S/N and resolution
factor).  Since PSF ellipticity is not a dominant driver of the object S/N or
resolution, this mismatch between data and simulations is acceptable.

\subsection{Shape measurement error}\label{subsec:sigmae}

In order to apply optimal weighting to the galaxy shapes when inferring ensemble shear, we need to
understand the shape measurement errors $\sigma_e$ as a function of galaxy properties.  The re-Gaussianization
algorithm includes a naive formula for shape measurement error, but this formula has been shown to
be incorrect by at least tens of percent in SDSS
\citep{2012MNRAS.425.2610R}, and does not account for correlated noise as in the HSC coadds, 
so we use the simulations in this paper to make a more accurate
estimate that will enable more optimal weighting.

The method adopted here relies on the fact that the simulations have pairs of galaxies that are
rotated by $90^\circ$ with respect to each other, but with nearly independent\footnote{They are
  `nearly' but not completely independent because noise due to light from 
  real objects on the outskirts of the galaxies is correlated between the original and rotated
  images.} realizations of the noise field.  The mathematical basis of our $\sigma_e$ estimates is
given in Appendix~\ref{app:sigmae}.

Our expectation is that to lowest order, the shape measurement error will depend on the detection
signal-to-noise ratio S/N and the resolution factor $R_2$.  For that reason, we define a sliding
window in that 2D parameter space, and estimate $\sigma_e$ as a function of position in that plane.
We also confirm that the 2D relationship defined in this way, $\sigma_e(\text{S/N}, R_2)$, does not
depend significantly on other parameters such as cuts on the magnitude or observing conditions
(which primarily shift around the distribution of the galaxies in that 2D plane without modifying
the $\sigma_e(\text{S/N}, R_2)$ relation).

\subsection{Intrinsic shape dispersion}\label{subsec:erms}
The per-component intrinsic shape dispersion, $e_\text{rms}$, is generally estimated from the shapes
in the real data with galaxies indexed $i$ as follows:
\begin{equation}\label{eq:ermsest}
e_\text{rms} = \sqrt{\frac{\sum_i w_i \left(e_{i,1}^2 + e_{i,2}^2 - 2 \sigma_e^2(\text{S/N}_i,
      R_{2,i})\right)}{2 \sum_i w_i}}.
\end{equation}
This is effectively a weighted shape dispersion with the measurement error subtracted off in
quadrature.  The main assumptions behind this equation are (1) that the measurement errors are
correctly estimated, and (2) that the measurement errors are independent of the shape noise.  We
rely on the calculation described in Section~\ref{subsec:sigmae} to ensure the reliability of the
first assumption.  For (2), while the magnitude of the measurement errors have a very slight
dependence on the galaxy ellipticity, the correlation between those errors is too mild to invalidate
Equation~\eqref{eq:ermsest}.  This equation is valid in the limit that the observed shape dispersion
can be modeled as the convolution of the intrinsic shape dispersion and that due to measurement
error; but no assumption is made about their functional forms.  \rev{We have explicitly tested its
  validity using simulations with parametric galaxy models on a grid of S/N and resolution factor
  values, and found that for S/N$\sim$20, the above assumptions are valid for galaxies with $|e|$
  values below the 90th percentile, while the remaining galaxies exhibit low-level anticorrelations between
  intrinsic shape and measurement error.  For S/N$>50$ it is valid for essentially the entire
  sample.  Based on our process of forward-modeling this effect and testing the validity of
  Eq.~\ref{eq:ermsest}, we estimate that the failure of its assumptions leads to a $\sim 2$\% bias
  in the estimated $e_\text{RMS}$ for objects with S/N$\sim$20, below which there are very few
  galaxies in our sample.  Given this low bias that is reduced rapidly at higher S/N, along with the
fact that the resulting shear biases (due to incorrect shear responsivity) will be directly
calibrated out using the simulations, we rely on Eq.~\ref{eq:ermsest} for our analysis despite its limitations.}

The weights \rev{in Equation~\ref{eq:ermsest}} are defined in
Eq.~\eqref{eq:wi}. In principle this is circular, since $e_\text{rms}$ appears in the weights.
To address this problem, we use an estimate from SDSS for $e_\text{rms}$ in the weights
\citep{2012MNRAS.425.2610R}, estimate $e_\text{rms}$ using those approximate weights, then use that
$e_\text{rms}$ estimate to update the weights, and confirm that if we iteratively try to re-estimate
$e_\text{rms}$ there are no significant changes.

As for $\sigma_e$, we estimate the intrinsic shape dispersion as a function of S/N and $R_2$.


\subsection{Shear calibration and additive bias}\label{subsec:calib}

The detailed process for estimating shears is given in appendix~3 of \citet{2018PASJ...70S..25M}.
Here we summarize the key equations, where the relevant quantities for shear estimation are as follows:
\begin{itemize}
\item Distortions $e_1$, $e_2$ in pixel coordinates.
\item Shape weights $w$:  These are defined as inverse variance weights summing in quadrature the
  measurement errors (Section~\ref{subsec:sigmae}) and intrinsic shape dispersion
  (Section~\ref{subsec:erms}).
\item Intrinsic shape dispersion per component $e_\text{rms}$: These are defined as in
  Section~\ref{subsec:erms}.
\item Multiplicative bias $m$ averaged over components.  These biases are determined through the
  procedure described in this section.  Thus, for the initial calculation in this section, we set
  $m=0$.  When testing the efficacy of our calibration bias corrections, we later use the $m$ values
  determined from the simulations.
\item Additive biases $c_1$, $c_2$: These are set to 0 for the initial calculation, and
  then determined from the simulations as described below.
\end{itemize}

The ensemble shear per component can be determined using all galaxies indexed $i$ in a given
subfield $j$.  The ensemble average distortion is an
estimator for the shear $g$,
\beq
(\hat{g}_\mathrm{1},\hat{g}_\mathrm{2}) = \frac{1}{2\cal R}
\langle(e_\mathrm{1},e_\mathrm{2})\rangle,
\eeq
where the `shear responsivity' ${\cal R}$
represents the response of the distortion to a small
shear \citep{1995ApJ...449..460K, 2002AJ....123..583B}; ${\cal R} \approx
1-e_\mathrm{rms}^2$, where $e_{\mathrm{rms}}$ is the RMS intrinsic
distortion per component.  Hence we begin by empirically estimating the responsivity for the source
population in that subfield as
\begin{equation}\label{eq:calr}
{\cal R}_j = 1 - \frac{\sum_i w_i e_{\text{rms},i}^2}{\sum_i w_i}.
\end{equation}
The lensing weights $w_i$ are defined as
\beq\label{eq:wi}
w_i = \frac{1}{\sigma_{e,i}^2 + e_{\text{rms},i}^2}
\eeq
where $\sigma_{e,i}$ is the estimate of the per-component shape measurement error due to pixel noise
for object $i$ (with a strong dependence on object properties such as S/N) and $e_{\text{rms},i}$ is
the per-component RMS intrinsic shape dispersion.  Note that in general that quantity is defined for
the ensemble; the $i$ is meant to indicate that we determine and model its slow variation with object
properties.

We can use the catalog estimates of calibration bias $m_i$ to derive an ensemble estimate for the
calibration bias for our sample:
\begin{equation}\label{eq:hatm}
\hat{m}_j = \frac{\sum_i w_i m_i}{\sum_i w_i}
\end{equation}
In the initial calculation, $\hat{m}_j=0$; it is only nonzero after we have determined an initial set
of calibration corrections.  Likewise, the additive correction term can be derived from a weighted
sum in each component $k$,
\begin{equation}\label{eq:hatc}
\hat{c}_{j,k} = \frac{\sum_i w_i c_{i,j,k}}{\sum_i w_i},
\end{equation}
which will be zero in our initial shear estimation.

Finally, these can be combined with the standard average over tangential shear estimates to get the
stacked shear estimator per component $k$
\begin{equation}\label{eq:hatgamma}
\hat{g}_{j,k} = \frac{\sum_i w_i e_{i,k}}{2 {\cal R}_j (1+\hat{m}_j)\sum_i w_i} - \frac{\hat{c}_{j,k}}{1+\hat{m}_j}.
\end{equation}

At the end of this process, we have the following quantities for each subfield $j$:
\begin{enumerate}
\item Estimated shear $\hat{g}_{j,k}$ per component.
\item True shear $g_{j,k}$ per component.
\item PSF shear $g_{j,k}^\text{(PSF)}$.
\item A weight for each subfield, $w_j$.  For this purpose, we use the weights that are defined to
  force the simulations to match the observing conditions in the data as defined in
  Section~\ref{subsec:reweights16a}, but do not do any inverse variance weighting since the
  uncertainties in the ensemble shear estimates do not vary strongly by subfield.
\end{enumerate}

Our model for the per-object systematics is that a given object has a property-dependent
multiplicative bias $m_{i,k}$ and a fraction of the additive PSF anisotropy that leaks into the ensemble
shear estimates, $a_{i,k}$, giving an additive term that depends on the galaxy properties and the PSF
that convolves it, $c_{i,j,k}\equiv a_{i,k} g_{j,k}^\text{(PSF)}$.
As for $\sigma_e$, we define a
sliding window in S/N and $R_2$ values, and then carry out a
weighted linear regression to the shear estimates averaged within bins in that 2D plane to
estimate $m_{i,k}$ and $a_{i,k}$:
\begin{equation}\label{eq:sysmodel}
\hat{g}_{j,k} - g_{j,k} = m_{i,k} g_{j,k} + a_{i,k} g_{j,k}^\text{(PSF)}.
\end{equation}
 We
then take the mean of the two shear components to estimate the $m_i$ and $a_i$ values that are 
relevant for a real lensing analysis (where the shear components are measured in the
tangential/radial coordinate system, effectively averaging over the components in the pixel basis).

Finally, to confirm the efficacy of our calibration corrections in our shear estimator in 
Equation~\eqref{eq:hatgamma}, we carry out a round-trip exercise wherein shear estimation is done in
the simulations using the simulation-defined $m_i$ and $c_{i,k}$ values based on each object's 
S/N and $R_2$ values.  While this should trivially provide us with unbiased shear estimates, we can
split the simulations by observational conditions and other properties such as the apparent
magnitude or photometric redshift, and confirm that the calibration corrections are still suitable
for the kinds of subsamples we encounter when measuring shear in the real data.

\subsection{Selection bias}\label{subsec:sel}

Selection bias is an effect wherein quantities used to select or weight the galaxies entering the
lensing analysis depend on the galaxy shape, and therefore the probability of a galaxy entering the
sample (or the weight assigned to a galaxy in the sample) depends on its alignment with respect to
the shear direction or the PSF anisotropy direction.  This 
results in a violation of the assumption that galaxy intrinsic
shapes are isotropically oriented.  If the selection probability or weight depends on the shear, there
will be a multiplicative bias, whereas if they depend on the PSF shape, there will be an additive
bias.

For this discussion, we will refer generically to quantities $X$ used for selection or weighting; as
described in \citet{2018PASJ...70S..25M}, the primary quantities used for selection\footnote{\rev{At
  earlier stages of running the HSC pipeline, additional selection criteria are imposed to create
  the shear catalog; the full list is given in \citet{2018PASJ...70S..25M}.  We do not separately
  investigate these earlier flag cuts because they are aimed at removing junk while leaving
  essentially all galaxies, and hence cannot cause a selection bias in the galaxy sample.  For
  example, figure 17 of \citet{2018PASJ...70S...5B} shows that the star/galaxy separation criterion
gives a galaxy sample that is essentially 100 per cent complete to the depth of the shear catalog.
In contrast, the resolution and magnitude cuts clearly remove a non-negligible number of real
galaxies -- potentially tens of per cent of a sample that might be defined with more liberal
selection criteria -- and hence can more easily result in selection bias in the resulting galaxy sample.}} are $R_2$,
$i$-band \texttt{cmodel} magnitude, and $i$-band \texttt{cmodel} S/N.  In addition, there are the
weights defined in Eq.~\eqref{eq:wi}.  Here we treat multiplicative selection biases as causing the
observed value of $\hat{X}$ (used for selection or weighting) to deviate from the true one
$\hat{X}$ as
\begin{equation}\label{eq:dx}
\Delta X \equiv \hat{X}-X \propto |g| \cos{(2\phi)},
\end{equation}
where $|g|$ is the lensing shear magnitude, and $\phi$ is the angle between the galaxy intrinsic
shape and the lensing shear.  The fact that this is proportional to $|g|$ can be thought of as a
Taylor expansion of some potentially more complex function, taken to the $|g|\ll 1$ regime.  The
$\cos{(2\phi)}$ term comes from basic symmetry considerations.  This equation is a generalization
of an expression derived for $\Delta R_2$ based on the Gaussian approximation in
\cite{2005MNRAS.361.1287M}, in which case the constant of proportionality is uniquely defined and
depends on the magnitude of the galaxy intrinsic distortion.  However, for the
\texttt{cmodel}-related quantities and the weights, which are all derived through complex multi-step
analyses, there is no obvious way to derive the constant of proportionality (or even what it depends
on) from first principles, so we rely on the simulations to determine this information. 
The sign of the constant of proportionality determines the sign of the shear bias.
A similar equation to Eq.~\eqref{eq:dx} can be written for additive selection biases, using the
magnitude of the PSF anisotropy and the angle between the galaxy and PSF shapes.

Below we describe separately how we quantify the edge-related and weight-related types of selection
effects so that we can remove the resulting biases in the ensemble shear signals.

\subsubsection{Selection bias due to weights}

If the weights in Eq.~\eqref{eq:wi} depend on galaxy shape, then even away from the boundaries of the
sample, we may get some selection bias due to our use of weighted ensemble averages (referred to as
`weight bias' by \citealt{2017MNRAS.467.1627F}).  Fortunately, the shear estimation process in the
simulations that was outlined in Section~\ref{subsec:calib} provides us with a way to estimate this
directly.  As noted there, when exploiting shape noise cancellation, we only use galaxies appearing
in 90$^\circ$ rotated pairs (applying our selection criteria at random to one galaxy in the pair),
and we enforce the same weight for galaxies in a pair.  However, if we  permit the two galaxies in
the pair to have different weights as determined by their individual values of S/N and $R_2$, then
the analysis will have weight bias built into it.  We can therefore estimate $m_\text{wt}$ and
$a_\text{wt}$, the multiplicative and additive bias parameters due to weight bias, empirically from
the simulations.

\subsubsection{Selection bias due to cuts}


For galaxies near the edges of our sample, the primary concerns are multiplicative and additive
biases due to the quantities used for cuts, $\hat{X}$, depending on the shear as in
Eq.~\eqref{eq:dx} or on the magnitude of the PSF anisotropy.  Conceptually, the magnitude of the
selection bias that we expect depends on two things: the prefactor in front of Eq.~\eqref{eq:dx},
which quantifies the strength of the dependence of $\hat{X}$ on the shear or PSF ellipticity; and
the fraction of the sample that is close enough to the edge that shear or PSF anisotropy can lead to
some coherent $\Delta \langle e_+\rangle$ along the shear or PSF anisotropy direction.  For a single
quantity $\hat{X}$, the relevant expression for $\Delta \langle e_+\rangle$ is
\citep[][equation 18]{2005MNRAS.361.1287M}
\begin{equation}\label{eq:dep}
\Delta \langle e_+\rangle = \int_{0}^{1} \mathrm{d}|e| \frac{\mathrm{d}^2 n}{\mathrm{d}|e|\,\mathrm{d}\hat{X}}\big{|}_\text{edge} \int_{0}^{\pi} \frac{\mathrm{d}\phi}{\pi} |e|\cos{(2\phi)}\Delta X.
\end{equation}
This equation implicitly assumes that we are placing a lower limit on $\hat{X}$; the sign flips for an
upper limit.   The
$\Delta X$ will in general be described by some equation like Eq.~\eqref{eq:dx} that cannot be
derived analytically, giving $\Delta X = A(|e|, X_\text{edge}) |g| \cos{(2\phi)}$. In this equation, $\frac{\mathrm{d}^2
  n}{\mathrm{d}|e|\,\mathrm{d}\hat{X}}\big{|}_\text{edge}$ is the joint distribution of distortion
magnitude and $\hat{X}$ evaluated at the edge of the sample in $\hat{X}$. For some $\hat{X}$ the
joint distributions of distortion magnitude $|e|$ and $\hat{X}$ are effectively independent, so
\begin{equation}
\frac{\mathrm{d}^2 n}{\mathrm{d}|e|\,\mathrm{d}\hat{X}}\big{|}_\text{edge} \equiv p(|e|) p(\hat{X})_\text{edge}.
\end{equation}
Simplifying Eq.~\eqref{eq:dep}, we find
\begin{align}
\Delta \langle e_+\rangle &= |g| p(\hat{X})_\text{edge} \int_{0}^{1} \mathrm{d}|e| \, |e| p(|e|)
A(|e|, X_\text{edge}) \times\notag\\ 
 & \qquad\qquad \qquad\qquad \qquad\qquad  \int_{0}^{\pi}
\frac{\mathrm{d}\phi}{\pi} \cos^2{(2\phi)}\\
&= \frac{1}{2}|g| p(\hat{X})_\text{edge} \int_{0}^{1} \mathrm{d}|e| \, |e| p(|e|) A(|e|, X_\text{edge}).
\end{align}
Since the added shear  due to a cut on $\hat{X}$ is proportional to the shear itself, we can write
this as a multiplicative shear selection bias that depends on the distribution of $\hat{X}$ as
\begin{equation}\label{eq:msel}
m_\text{sel}(\hat{X}) \propto  p(\hat{X})_\text{edge}.
\end{equation}
We are absorbing the complexity of how $\Delta X$ depends on the galaxy shape and other properties
in the constant of proportionality.
By a similar formalism,
\begin{equation}\label{eq:asel}
a_\text{sel}(\hat{X}) \propto  p(\hat{X})_\text{edge}.
\end{equation}
Our approach is to assert these relationships, estimate the factors of proportionality directly in the
simulations, and then test whether those factors of proportionality still hold if we change the cuts
we apply.  In order to estimate the factors of proportionality directly in the simulations, we
estimate the shear {\em without} imposing shape noise cancellation.  In this case, the two
galaxies in the $90^\circ$ rotated pairs are selected based on their own $\hat{X}$ values, and both
additive and multiplicative selection biases should be evident.

Because of our expectation that the selection biases should be directly proportional to
$p(\hat{X})_\text{edge}$ for each cut quantity $\hat{X}$, we focus on resolution factor
and magnitude.  As shown in Fig.~\ref{fig:data-sims-overall}, these two cuts affect the sample quite
significantly, so the distribution of both quantities is non-negligible near the edge of the
sample.  In contrast, the S/N cut removes less than one per cent of the objects passing the other
two cuts.  Additional cuts given in \citet{2018PASJ...70S..25M}, such as on blendedness and
multiband S/N, function primarily as junk rejection cuts that remove tiny fractions of the
sample, and hence we do not carry out this selection bias analysis for them.

\subsection{Outputs}

Using the analyses described earlier in this Section, the simulations are used to provide the
following catalog entries for use in real lensing analysis:
\begin{itemize}
\item Shape weights {\tt ishape\_hsm\_regauss\_derived\_weight} based on the inverse variance
  (Section~\ref{subsec:sigmae} and~\ref{subsec:erms}).
\item Intrinsic shape dispersion per component {\tt ishape\_hsm\_regauss\_derived\_rms\_e} (Section~\ref{subsec:erms}).
\item Multiplicative bias {\tt ishape\_hsm\_regauss\_derived\_bias\_m} (Sections~\ref{subsec:calib}
  and~\ref{subsec:sel}).
\item Additive biases {\tt ishape\_hsm\_regauss\_derived\_bias\_c1/c2} (Section~\ref{subsec:calib}).
\end{itemize}

In addition, there are ensemble corrections for selection bias due to the cuts placed on S/N, $R_2$, and
$i$-band magnitude (Section~\ref{subsec:sel}).

\section{Results}\label{sec:results}

In this section we describe the results of analyzing the simulations.  For the majority of the
results described here, we use parent sample 4 (Table~\ref{tab:parentsamp} and
Section~\ref{subsubsec:finalps}).  However, in Section~\ref{subsec:results-samples}, we will show
results from all parent samples to justify this choice and illustrate the impact of some of the
choices of parent sample definition. 


\begin{figure*}
\begin{center}
\includegraphics[width=0.9\textwidth]{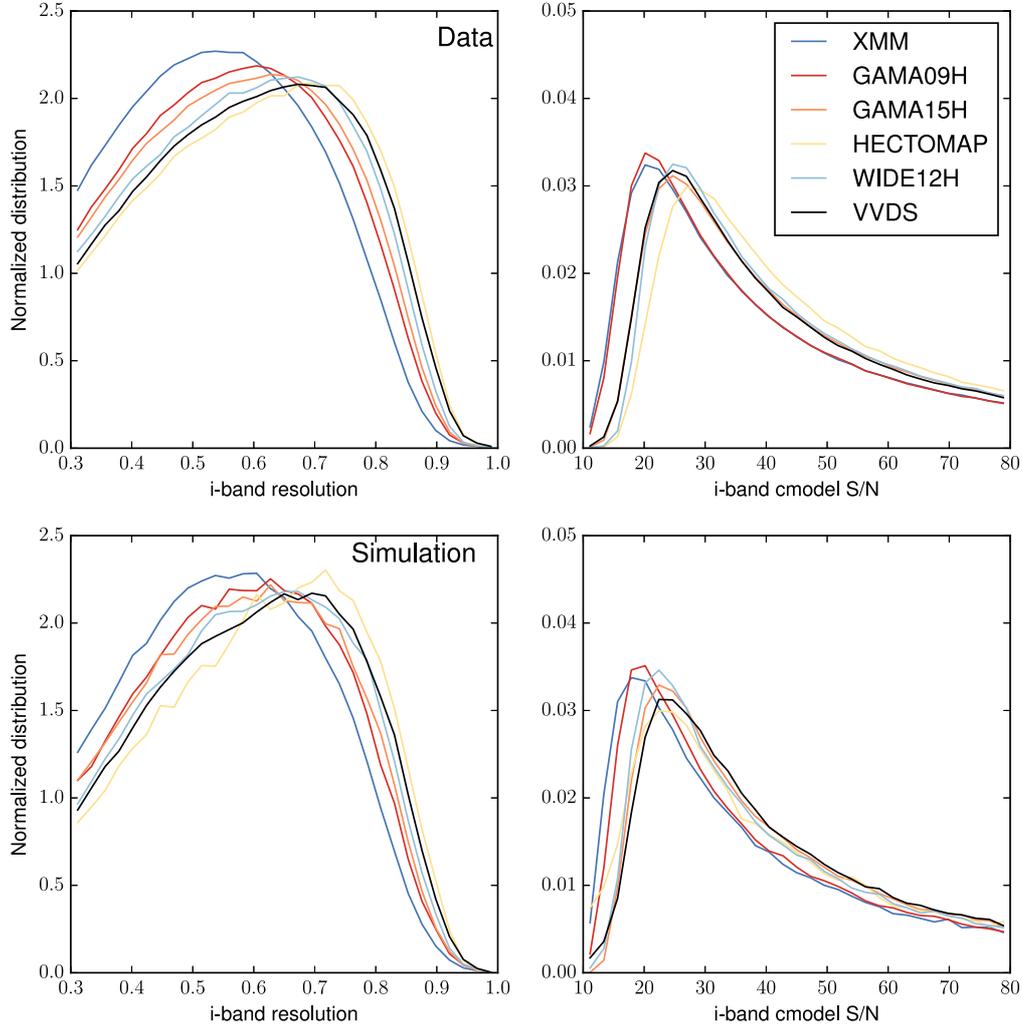}
\end{center}
\caption{Weighted distribution of $i$-band resolution factor (left) and $i$-band cmodel S/N (right)
  for the HSC S16A shear catalog (top) and the simulations (bottom), shown separately for each field.
As shown, the distributions of these two quantities exhibits significant field-dependence, due to
the field-dependent variation in observing conditions.  The simulations are largely able to
reproduce this field-dependence, especially for the resolution factor. This plot shows results for our default
simulations, which use parent sample 4 (Table~\ref{tab:parentsamp} and
Section~\ref{subsubsec:finalps}).  \label{fig:data-sims-byfield}}
\end{figure*}
\begin{figure*}
\begin{center}
\includegraphics[width=0.9\textwidth]{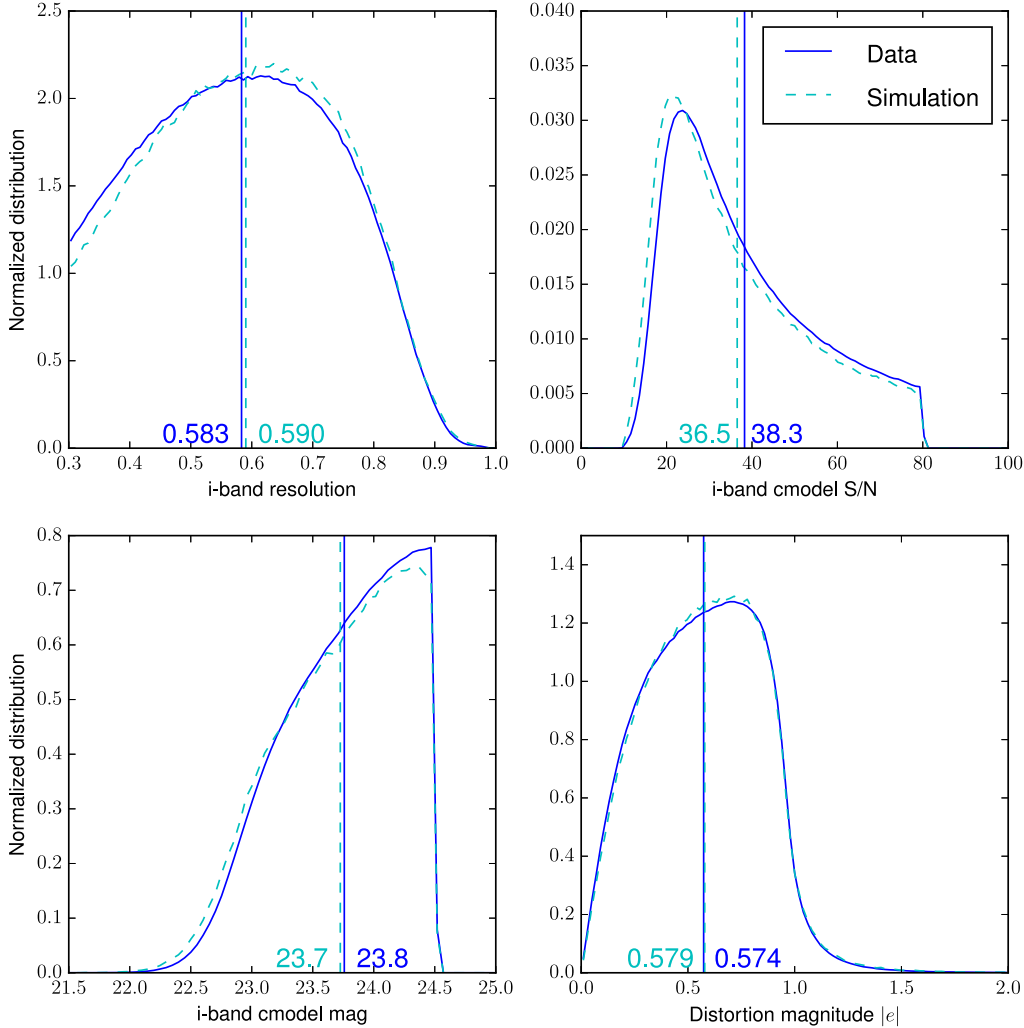}
\end{center}
\caption{Weighted distribution of $i$-band resolution factor (top left), $i$-band cmodel S/N (top
  right), $i$-band cmodel magnitude (bottom left), and distortion magnitude $|e|$ (bottom right),
  shown for the HSC S16A shear catalog and the simulations, as an average over the entire
  survey.  The weighted mean values are quoted and shown as vertical lines on the
  plots.  This plot shows results for our default
simulations with parent sample 4 (Table~\ref{tab:parentsamp} and
Section~\ref{subsubsec:finalps}), which reproduce the properties of galaxies in the simulations
quite well. \label{fig:data-sims-overall}}
\end{figure*}
\begin{figure*}
\begin{center}
\includegraphics[width=\textwidth]{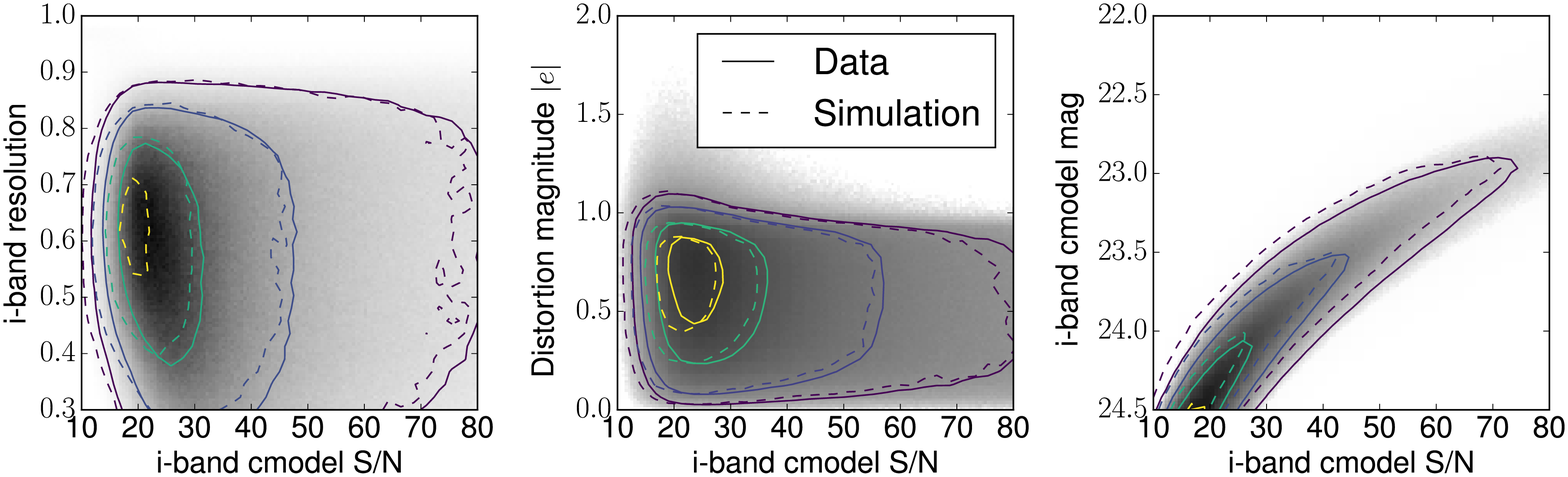}
\end{center}
\caption{The weighted 2D distributions of resolution factor (left), distortion magnitude $|e|$ (middle), and
  cmodel magnitude (right), vs.\ the $i$-band cmodel S/N.  The grayscale shows the 2D density in the
data, while the solid and dashed contours are for the data and simulations, respectively.  This plot shows results for our default
simulations, which use parent sample 4 (Table~\ref{tab:parentsamp} and
Section~\ref{subsubsec:finalps}). \label{fig:data-sims-2d}}
\end{figure*}

\subsection{Galaxy properties}

In this section, we compare the distribution of observable galaxy properties that can be inferred from
the images in the simulations and data.  The
first comparison is shown in Figure~\ref{fig:data-sims-byfield}.  Here, we compare the weighted
distributions of $i$-band resolution factor (left) and $i$-band cmodel S/N (right) for the HSC S16A
shear catalog (top) and the simulations (bottom).  The distributions have been computed separately
for each field, because of the substantial variations in observing conditions between these fields
shown in Figures~\ref{fig:noise-var} and~\ref{fig:psf-size} (however, results averaged across fields
are shown in Figure~\ref{fig:data-sims-overall}).  We expect that those differences in
observing conditions should lead to differences in the observed properties of galaxies passing our
selection criteria, and this figure confirms that that is the case.  The differences in PSF FWHM are
more striking than the differences in noise variance between the fields.  When ranking the fields
based on the mean PSF FWHM, they are (from best to worst) HECTOMAP, VVDS, WIDE12H, GAMA15H, GAMA09H,
and XMM.  As a result, we expect that the average galaxy resolution factor and S/N will be highest
for HECTOMAP and lowest for XMM.  The results are consistent with those expectations in both the
simulations and data.

In contrast, we find (not shown) that the distributions of $i$-band magnitude and distortion magnitude $|e|$
exhibit relatively little field-dependence.  The reason for this is that our magnitude cut of $24.5$
in $i$-band corresponds to a S/N that in most cases is substantially higher than the limits in
typical weak lensing surveys (and higher than our limit of $10$, which eliminates only a tiny
fraction of galaxies that would pass our other cuts).  Hence the magnitude distribution does not
depend strongly on observing conditions, but for a fixed magnitude, the S/N distribution does depend
strongly on observing conditions.

Now that we have confirmed that field-dependent trends in galaxy properties are consistent in the
data and simulations, we compare the distributions of galaxy properties across the survey overall in
simulations and data in Figure~\ref{fig:data-sims-overall}.  As shown, the best match is between the
distributions of $|e|$ and $i$-band cmodel magnitude; the means of those distributions agree between
the data and simulations to better than 0.1 per cent and 1 per cent, respectively.  The mean values
of resolution factor and $i$-band cmodel S/N values differ by 1 and 5 per cent when comparing data
vs.\ simulations, with some more visually obvious discrepancies in the shapes of the distributions.
One difficulty in matching the S/N distribution is that the variances in the data were rescaled
during the simulation process to approximately correct for the impact of noise correlations on S/N
estimates.  We have attempted to mimick this rescaling in the simulations, but did not have all the
information needed to ensure the right range of rescaling factors as was actually used in the data
for the weak lensing catalog. 
Considering that we selected these galaxy samples from real data and then did a full simulation
process including blends and other complications, this level of agreement is impressive. 
For a comparison with the state of the art in other surveys, see figure~3 in
\cite{2017MNRAS.467.1627F} from the Kilo-Degree survey.  There, the weighted distributions of S/N,
magnitude, scale length, and $|e|$ in data vs.\ simulations show visually obvious discrepancies that
appear to exceed the level achieved here for size, shape, and magnitude.
For the Dark Energy Survey Year 1 shear catalogs, \citet{2017arXiv170801533Z} figure 12 shows a
comparison of galaxy properties in the simulations used to calibrate one set of shear estimates
against the data.  The level of agreement exhibited there is comparable to that in our
Figure~\ref{fig:data-sims-overall}, with low-level tension between the distributions of size and S/N
in simulations and data, and excellent agreement on the distribution of galaxy shapes.

Finally, in Figure~\ref{fig:data-sims-2d} we confirm that the shapes of the 2D joint distributions
are well-matched between simulations and data, in addition to the shapes of the 1D distributions
shown in previous figures.


\subsection{Shape measurement uncertainty}\label{subsec:results-sigmae}

\begin{figure}
\begin{center}
\includegraphics[width=\columnwidth]{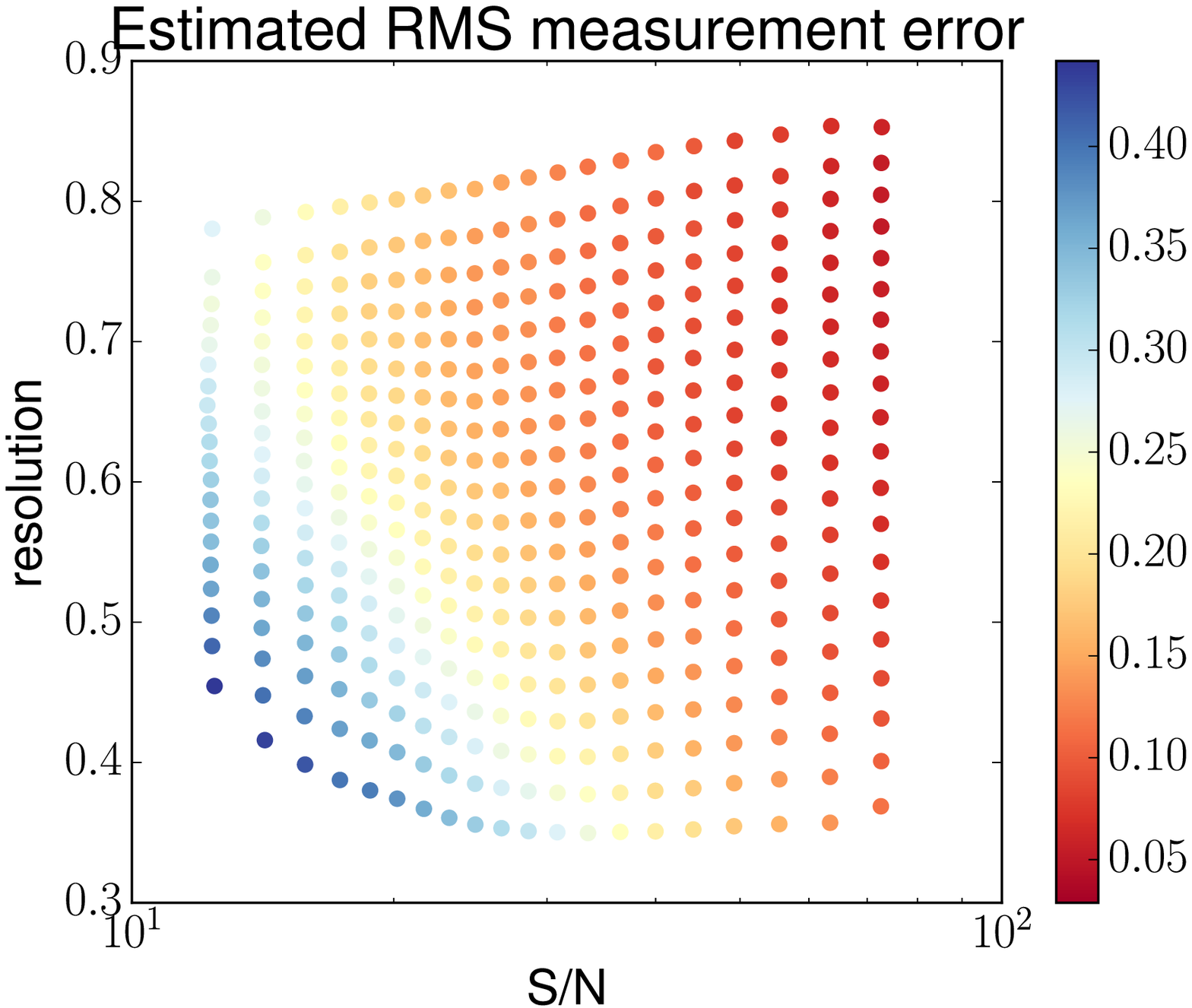}
\includegraphics[width=\columnwidth]{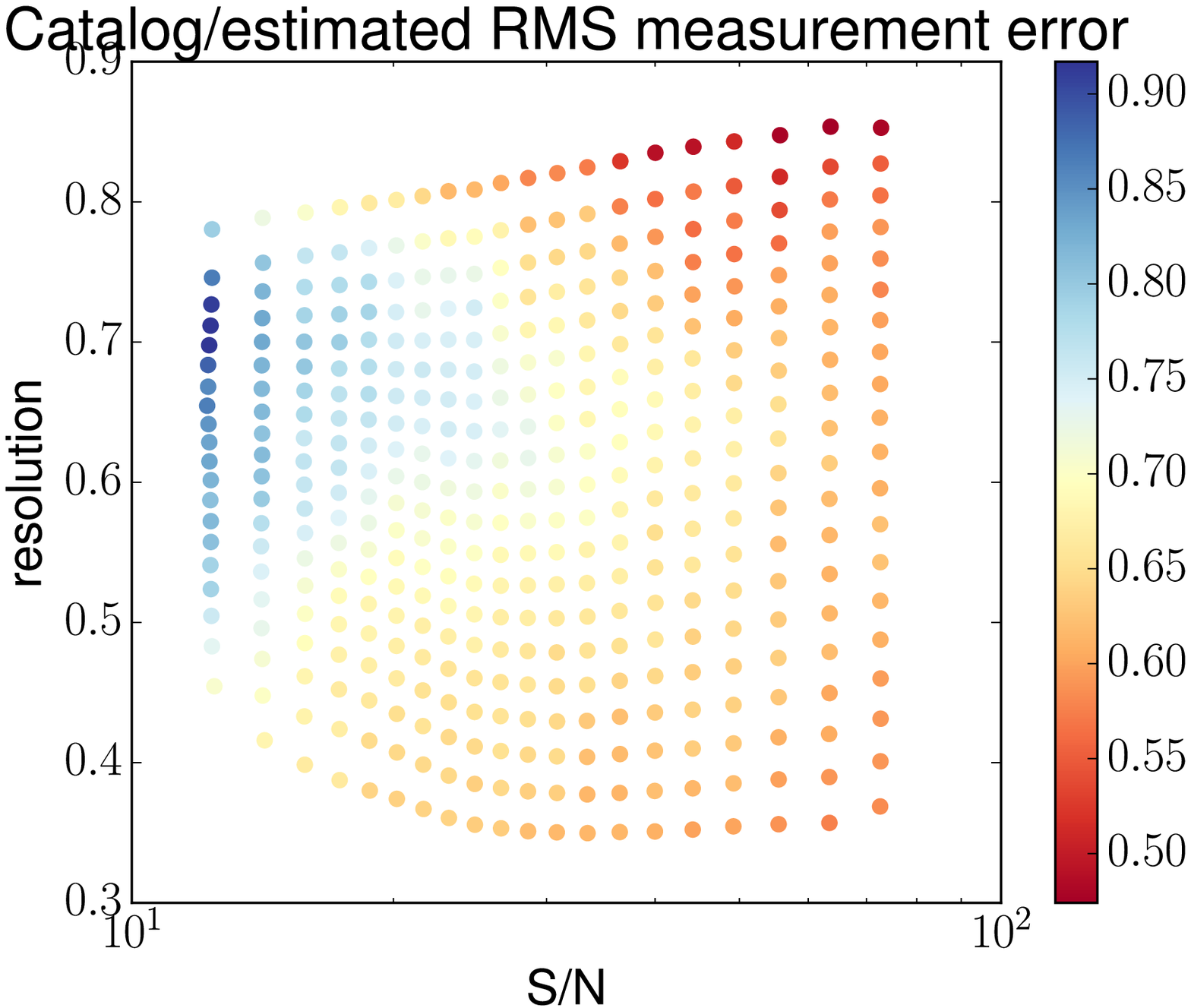}
\end{center}
\caption{The top panel shows the estimated per-component RMS distortion, $\sigma_e$, from the
  simulations as a function of the $i$-band cmodel S/N (horizontal axis) and the $i$-band
  resolution factor (vertical axis).  The bottom panel shows the ratio of the catalog values of
  $\sigma_e$ to those estimated from the simulations, illustrating the fact that the catalog
  estimates are consistently too low compared to reality.\label{fig:sigmae}}
\end{figure}

Following the procedure from Section~\ref{subsec:sigmae}, we determined the per-component shape
measurement uncertainty as a function of S/N and $R_2$.  This function is shown in
Figure~\ref{fig:sigmae}, where the top panel shows the estimated value and the bottom panel shows the
ratio of the values initially estimated by the HSC pipeline to those estimated from the
simulations.  The HSC pipeline values underestimate the measurement error in the distortion for all galaxies, with a
typical 30 per cent underestimation but with some dependence on object properties.

In order to estimate $\sigma_e$ for each galaxy in our catalog, we need a method of
interpolating in this 2D plane.  Given the steep dependences of $\sigma_e$ on these properties, we
do not directly interpolate the function shown in Fig.~\ref{fig:sigmae}. Instead, we fit for a
power-law $\sigma_e(\text{S/N}, R_2)$ relation:
\begin{equation}
\sigma_e = 0.264 \left(\frac{S/N}{20} \right)^{-0.891} \left(\frac{R_2}{0.5} \right)^{-1.015}.
\end{equation}
Then, we linearly interpolate the ratio of the
estimated $\sigma_e(\text{S/N}, R_2)$ to that power-law based on the $\log_{10}{(\text{S/N})}$ and
$R_2$ values, and use it to correct the power-law model.  This means we are interpolating a function
that is much closer to constant, varying by at most 25 per cent (and for typical parameter values,
10 per cent).  For S/N and $R_2$ values outside
the bounds of our sliding window, we simply use the nearest point within our sliding window for the
interpolation of this ratio, but we still use the exact value of the power-law for the baseline
$\sigma_e(\text{S/N}, R_2)$ model.  


\subsection{Implications for the RMS distortion}
\begin{figure}
\begin{center}
\includegraphics[width=\columnwidth]{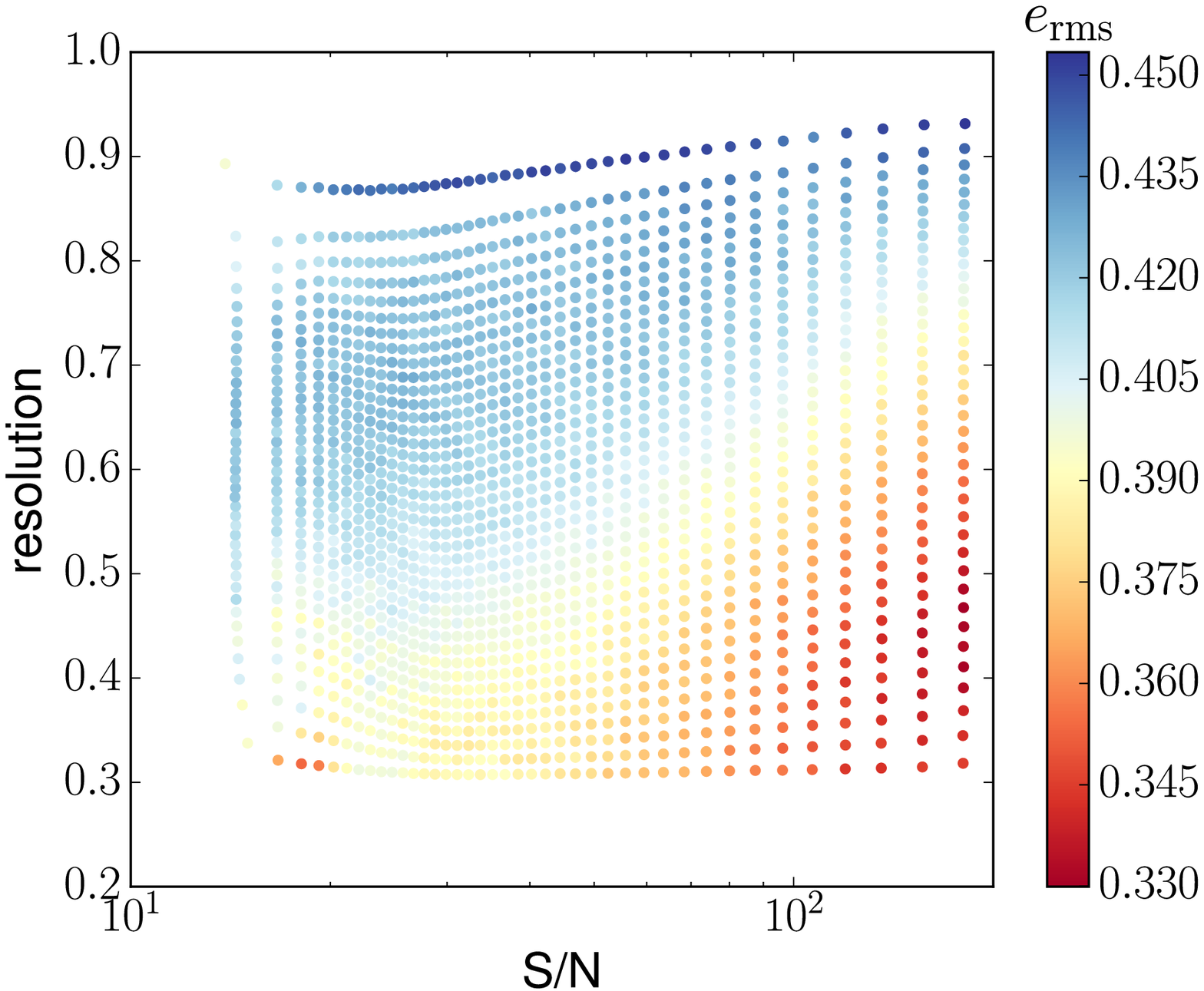}
\end{center}
\caption{The per-component intrinsic shape dispersion estimated from the HSC shear catalog after subtracting the
  measurement error estimated from the simulations.  For the majority of the galaxy sample
  (S/N$\lesssim 80$), the intrinsic shape dispersion is relatively flat as a function of object
  properties.
\label{fig:intrinsic_e}}
\end{figure}

Following the process described in Section~\ref{subsec:erms}, we use the measurement noise function
estimated in Section~\ref{subsec:results-sigmae} to estimate the intrinsic shape dispersion in the
real HSC shear catalogue.  In other words, we use the {\em real} data, estimate the observed shape
dispersion per component, and subtract off (in quadrature) the contribution from measurement noise
based on the simulation-calibrated $\sigma_e$ values.  The resulting intrinsic shape dispersion is
shown in Figure~\ref{fig:intrinsic_e}. It is possible that the rise in the estimated dispersion at
large $R_2$ is because that part of parameter space has a
relatively higher fraction of unrecognized blends.

As shown, this function is relatively flat, with a value around 0.4 for most of parameter space.
This corresponds to $\sigma_\gamma\approx \sigma_e/(2 (1-0.4^2))\sim 0.24$.  Each object in the
catalog has a value of $e_\text{rms}$ linearly interpolated according to its value of resolution
factor and log(S/N).  Together with the $\sigma_e$ values from Section~\ref{subsec:results-sigmae},
these $e_\text{rms}$ uniquely determine the optimal (inverse variance) shape weighting given in
Eq.~\eqref{eq:wi}.

\rev{Because of the way that we have estimated the intrinsic shape dispersion in the real data, by
  design it must be the case that $\sqrt{\langle\sigma_e^2+e_\text{rms}^2\rangle} =
  \sqrt{\langle 0.5(e_1^2+e_2^2)\rangle} $ in bins of
  S/N and resolution factor in the real data.  This is definitionally
  the case; however, a non-trivial test for whether we are missing dependencies on other parameters
  that should have been in our model is to test whether that equality holds when splitting the
  sample by other parameters.  The reason this test is valuable is that in real data, we may split
  the sample by some other parameters, and we need to be able to use the model for $e_\text{rms}$ in
order to infer the correct shear responsivity and hence the ensemble shear.}

\rev{When using i-band magnitude as an additional parameter, we find that this equality holds to within 1
  per cent for the entire magnitude range of our sample.  Since the shear responsivity is roughly
  $1-e_\text{rms}^2$, this implies that uncertainties in the shear responsivity (and calibration)
  when dividing the sources by magnitude are at most 0.4 per cent.  A more substantial concern
  arises when dividing the sample based on photometric redshift, which is a very common use case for
tomographic weak lensing analyses.  There, deviations from equality that are of order 8 per cent
were seen for photometric redshifts between 1--1.5 and $\text{S/N}>40$, resulting in potential shear calibration biases
of order 3 per cent at most (assuming that the entirety of this 8 per cent deviation is due to
misestimation of $e_\text{rms}$ and not $\sigma_e$).  Using our canonical parent sample of galaxies
(sample 4) we cannot directly disentangle the possible misestimates of $e_\text{rms}$ and $\sigma_e$
in this redshift range, because we center our cutouts at the
locations of HSC detections down to the detection limit without good redshift estimates.  Hence we
revisit parent sample 2 to understand this effect further.  For that parent sample, we have
high-quality 30-band photometric redshifts from COSMOS that can be used to reliably divide the
sample based on redshift.  Using those redshifts, we directly confirm that at fixed S/N and
resolution factor, $\sigma_e$ has almost no dependence on redshift, while $e_\text{rms}$ depends on
the joint distribution of redshift and S/N in a pattern consistent with what was seen in the real HSC data.  The effect is
such that for a sample in the photo-$z$ range $[1,1.5]$, there is a positive shear calibration bias
of $+3$ per cent.  Studies that divide the shear catalogue in ways that emphasize this redshift
range, and that have a high level of statistical precision (such that a 3 per cent bias would be
significant), should correct for this effect when modeling the shear signals.}

\subsection{Multiplicative shear calibration}\label{subsec:results-shearcalib}
To determine the baseline shear calibration in the absence of selection biases, we carry out the
shear estimation process and calibration bias determination based on the process in
Section~\ref{subsec:calib} while using shape noise cancellation.  Here, we always keep both objects
in any 90$^\circ$ rotated pair, imposing our weak lensing selection criteria on {\em only one
  randomly chosen object in the pair} and requiring them to both use the weight factor for that galaxy.  By doing so, we ensure that our selection does not lead to
some preference for the shear or PSF direction, enabling us to separate out basic features of the
PSF-correction method such
as model bias and noise bias from selection biases.

\begin{figure}
\begin{center}
\includegraphics[width=\columnwidth]{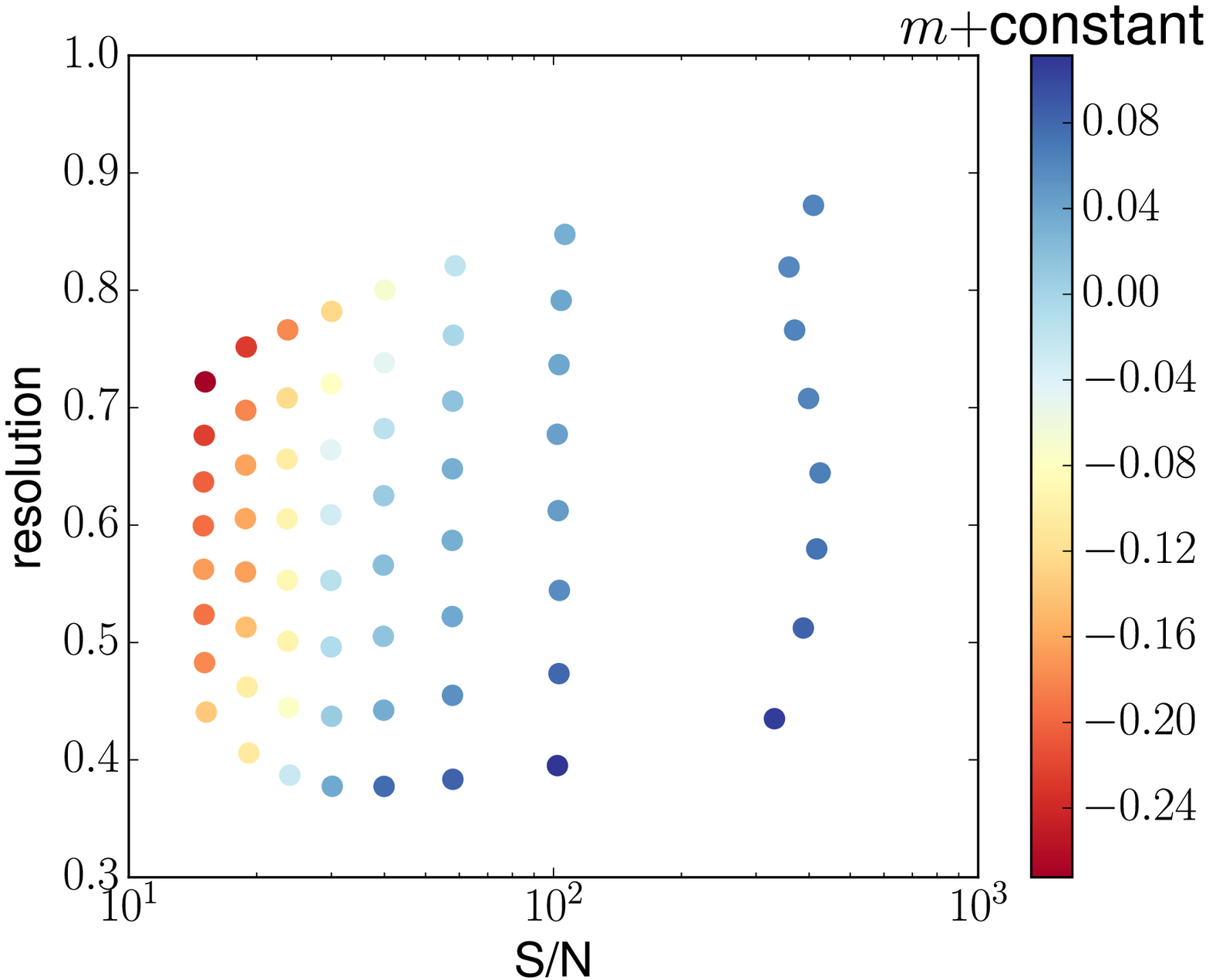}
\includegraphics[width=\columnwidth]{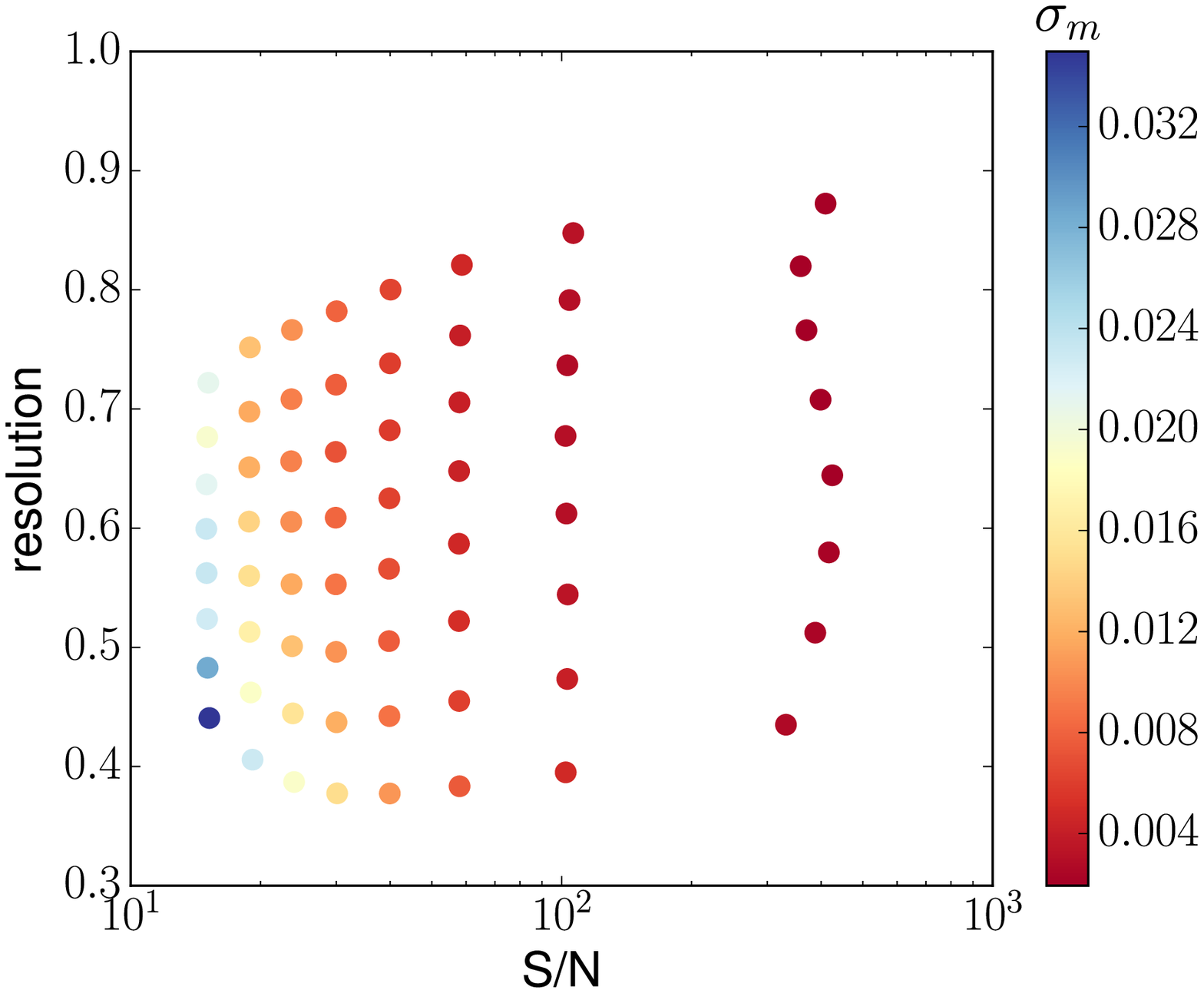}
\end{center}
\caption{{\em Top:} The component-averaged multiplicative shear bias $m$ in the 2D S/N vs.\
  resolution factor ($R_2$) plane, with an unspecified offset to preserve the blinding of
  the shear calibration. {\em Bottom:} The statistical
  uncertainty in the per-bin $m$ estimates.\label{fig:m-2d}}
\end{figure}

In Figure~\ref{fig:m-2d}, we show the calibration bias as a function of S/N and resolution factor
for the simulations overall, after subtracting off an unspecified constant value in order to
preserve the blinding of our shear analysis.  For reference, the bottom panel shows the statistical
error for each of the points in parameter space on the top panel.  Similarly to what was done for
shape measurement error in Section~\ref{subsec:results-sigmae}, we fit $m(\text{S/N}, R_2)$ to a
power-law in both parameters plus a constant additive offset, and interpolate a correction to that
power-law based on the difference between the values in Fig.~\ref{fig:m-2d} to the best-fitting
model.  This defines our baseline calibration bias correction, which will be distributed with the
public shear catalog.

For this baseline correction, the calibration bias is more negative at low S/N and high $R_2$.  The
power-law scaling with S/N is $m\propto (\text{S/N})^{-1.24}$.  The expected scaling with
S/N for a pure noise bias in the case of Gaussian galaxies and PSFs and moments-based shape
measurement would be $m\propto (\text{S/N})^{-2}$ \citep{2004MNRAS.353..529H,2012MNRAS.425.1951R},
steeper than our results.  However, it is not clear that the Gaussian approximation should be valid
in this case given the realistic complexity in galaxy morphologies and PSFs, nor is the \texttt{cmodel} S/N
estimate necessarily the one determining this relation.

The goal of the remainder of this section is to determine the validity and robustness of that
calibration bias correction, and assign a systematic uncertainty to it.  For example, we will
explore the validity of the assumption that we can correct for calibration bias as function of just
those two parameters, not including the magnitude or observing conditions explicitly.  Since
different lensing analyses will have different photometic redshift cuts (which correspond to complex
cuts in the multidimensional color-magnitude space) and different area cuts (e.g., if using a galaxy
cluster sample in a subset of the HSC survey region as lenses), this sensitivity analysis is
necessary to define the systematic error budget due to shape measurement errors in the case of
perfect PSF modeling.

In \citet{2018PASJ...70S..25M} equation~(6), a requirement on the systematic
uncertainty in the shear calibration bias is given: $|\delta m|$ is required to be below 0.017 in order to
avoid contaminating a cosmological galaxy-galaxy lensing analysis at the $0.5\sigma$ level,
integrated over all scales.  This is more stringent than the requirement derived for cosmic shear
analysis, and hence we adopt this more stringent requirement.  It is important to bear in mind two
points about this number.  First, calibration biases that are {\em well-understood} do not
contribute here.  If the biases are well-understood, we quantify and remove them using the
aforementioned baseline correction.  
Second, the requirement is defined such that in a cosmological weak lensing analysis with the
highest possible S/N for first year data would have systematic uncertainties due to shear
calibration bias that can be effectively ignored in the analysis.  Weak lensing analyses that
achieve lower integrated S/N, for example due to use of a different lens sample or range of physical
separations, can tolerate higher systematic errors than this requirement.  Moreover, even the
highest-precision measurements can still be done if the systematic errors exceed this requirement,
but in that case, the failure to meet this requirement would require explicit tracking of the
calibration bias uncertainty in the error budget, and the systematic uncertainty would no longer be
negligible compared to the statistical error.

To carry out the sensitivity analysis and ascertain whether we achieve our requirements on shear
estimation-related calibration bias, we proceed as follows:
\begin{itemize}
\item Define specially-selected subsamples of the shear catalog, e.g., with magnitude cuts, cuts on
  seeing, or other subsamples that do not correspond explicitly to selection in S/N and $R_2$.
\item Estimate the shear in the simulations for those subsamples, including the calibration bias
  corrections as per the shear estimator in Eq.~\eqref{eq:hatgamma}.
\item Use the systematics model in Eq.~\eqref{eq:sysmodel} to estimate the residual calibration bias
  for the subsample.
\end{itemize}
By definition, this `round-trip' exercise should give $m=a=0$ when using the entire weak lensing
sample in the simulations, but may not give a calibration bias consistent with zero if our
assumption that calibration bias depends only on S/N and $R_2$ is faulty.  The results of this
analysis are summarized in Table~\ref{tab:sensitivity}.

First, as shown in the table, our calibration bias corrections lead to an overall ensemble shear
bias of $-0.001\pm 0.003$. This is what is expected by default.

The next section of the table shows the results when splitting in quartiles based on the PSF FWHM.
This is meant to be an extreme case of what might happen when splitting up the survey based on
region; in general, area-related cuts do not lead to as extreme differences in seeing as a strict
seeing cut.  As shown, when dividing into quartiles in the seeing, the resulting calibration biases
exceed our requirements tolerances in two of the four cases.  The implications of this finding are
that weak lensing analyses that come with strict area cuts should evaluate the seeing distribution
after their cuts, and then use the simulations to evaluate whether there are additional shear
calibration biases that need to be removed for that special area definition.  The results of this
test are not relevant to the CMASS galaxy-galaxy lensing analysis that was used to set requirements,
because the CMASS sample covers the entire HSC survey area.

The third section of the table shows what happens if we change the limiting magnitude from 24.5 to
something lower (brighter).  Since the calibration bias corrections were defined in terms of S/N
instead of magnitude, any residual scalings with magnitude could result in a nonzero calibration
bias for the ensemble.  As shown, a cut at $i=24$, which eliminates tens of per cent of the sample,
does not lead to any statistically significant calibration bias ($0.002\pm 0.002$).  A cut at
$i=23.5$, which is a very severe limiting of the sample, gives a marginal ($2.5\sigma$) detection of
a calibration bias, $-0.005\pm 0.002$.  Nonetheless, this uncertainty is still well within our
requirement of $|\delta m|<0.017$, so this test does not raise any red flags for shear calibration
uncertainties in cosmological weak lensing analyses.


\begin{table}
  \caption{Sensitivity analysis for calibration bias correction $m$.  We present the ensemble average
    $m$ for subsamples of the simulated galaxy sample defined below, after imposing our baseline
    calibration corrections.}\label{tab:sensitivity}
\begin{center}
\begin{tabular}{ll}
\hline
Subsample definition & $m$ \\ \hline
Sample overall & $-0.001\pm 0.003$ \\ \hline
PSF FWHM: 1st quartile & $0.043\pm 0.005$ \\
PSF FWHM: 2nd quartile & $0.004\pm 0.005$ \\
PSF FWHM: 3rd quartile & $-0.019\pm 0.005$ \\
PSF FWHM: 4th quartile & $-0.011\pm 0.005$ \\ \hline
$i$ mag $<24$ & $0.002\pm 0.002$ \\
$i$ mag $<23.5$ & $-0.005\pm 0.002$ \\ \hline
\end{tabular}
\end{center}
\end{table}

Another potential contributor to systematic uncertainty in the shear calibration is failure of basic
assumptions about the parent sample in the simulation.  Since additional simulation sets are
required to explore this source of uncertainty, we defer this investigation to
Section~\ref{subsec:results-samples}.  As mentioned earlier in this section, we explicitly separate
out selection bias, and present estimates of its value and contribution to systematic error budget
in Section~\ref{subsec:results-selection}.

\rev{The final contributor to systematic uncertainty in the shear calibration that we consider here
  is the incorrect assumption in the simulations that the noise correlation function is spatially
  invariant across the HSC survey (see Section~\ref{subsec:noise}).  Our calculations described
  there indicate that the typical spatial scale on which the noise correlations varies is larger
  than the typical size of galaxies.  Hence, the primary impact of our simplified assumption is that
  a given galaxy that would be seen in the HSC survey with several different levels of noise
  correlation (if observed at different places on the sky) actually is simulated with only a single
  value of noise correlation.  As shown by \citet{2016MNRAS.457.3522G}, a given realization of that
  galaxy $j$ would have a different effective S/N$_j$ when changing the level of noise correlations
  even for fixed PSF and noise variance, and hence different levels of shear calibration bias
  (driven by noise bias) $m_j$ for the different realizations.  If $\langle m_j\rangle$ across those
  realizations is not equal to the estimated calibration bias $m$ at the single simulated level of
  noise correlations, this would mean that our shear calibration bias estimates are incorrect.}

\rev{Using multiple noise realizations of individual galaxy models, we have estimated how (for
  typical galaxies) the effective S/N changes when the PSF and noise variance are fixed, but the
  level of noise correlations varies within the level seen in individual patches in HSC.  We find
  that within that narrow range of variation, the S/N is linearly proportional to the level of noise
correlations at a separation of one pixel, and the range of S/N variation implied is $\pm 2$ per
cent from the S/N at the average level of noise correlations across the patch.  Since we have found
that $m\propto (\text{S/N})^{-1.24}$, this implies that the value of $m$ is changing by $\pm 2.5$
per cent of its value.  In other words, for a point in parameter space where we have estimated $m=-0.1$,
the variation in noise correlations across the survey means it really ranges from
$[-0.1025,-0.976]$, which is far below both the statistical uncertainty in $m$ and other sources of
systematic uncertainty in $m$.}

\subsection{Additive biases}\label{subsec:results-add}

In this section we show our results for the additive biases following the same methodology as for
multiplicative biases in Section~\ref{subsec:results-shearcalib}.

\begin{figure}
\begin{center}
\includegraphics[width=\columnwidth]{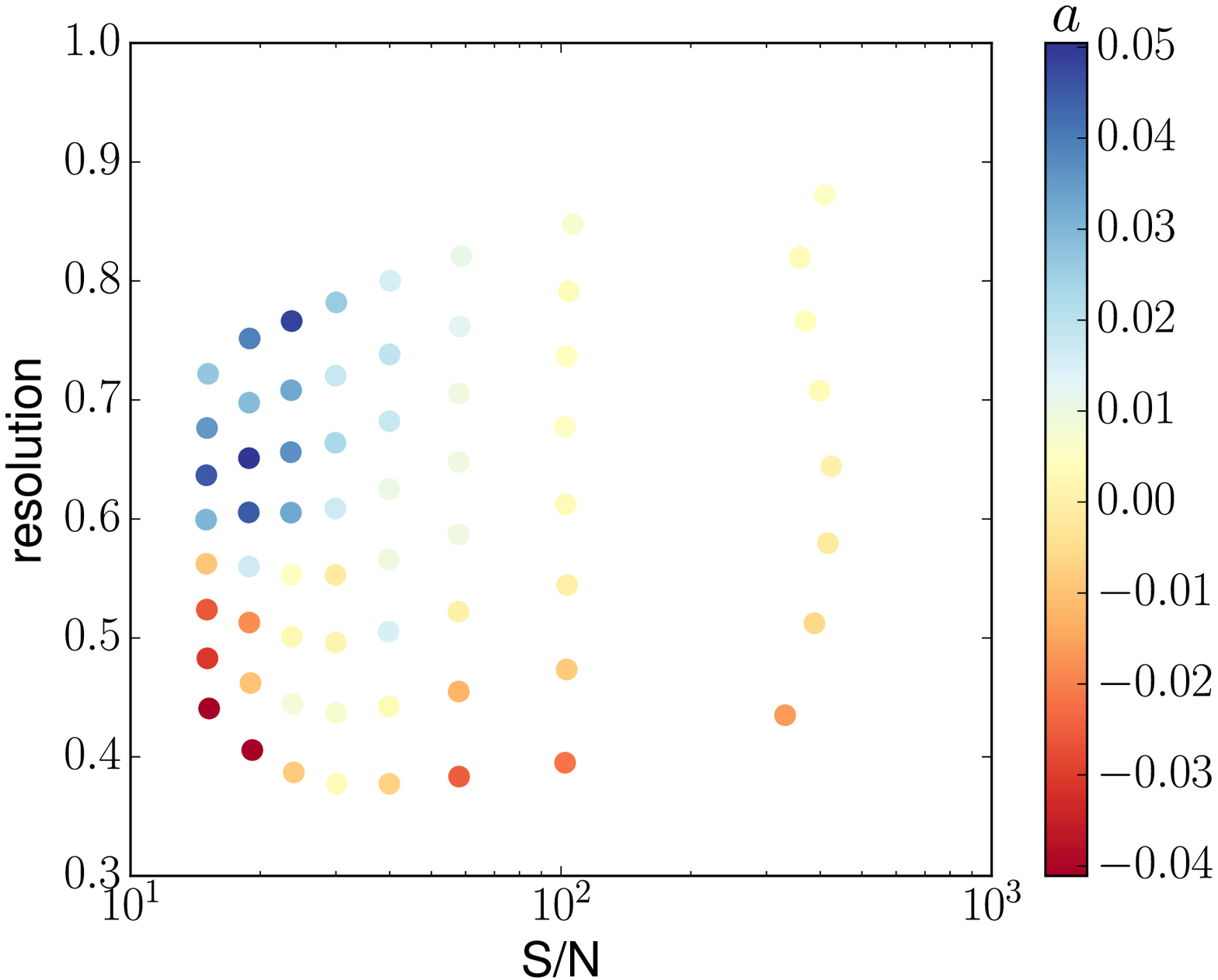}
\includegraphics[width=\columnwidth]{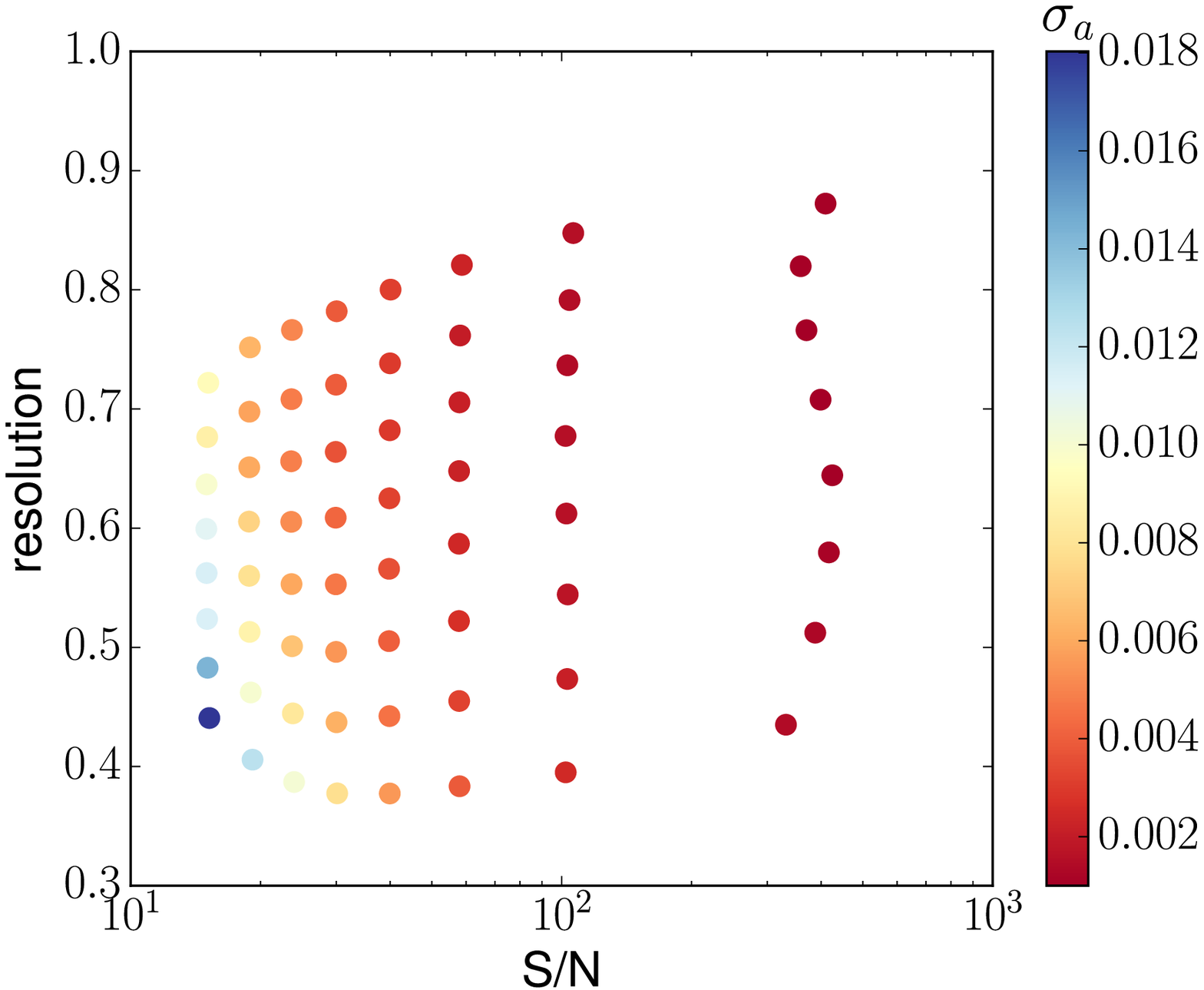}
\end{center}
\caption{{\em Top:} The component-averaged additive shear bias $a$ in the 2D S/N vs.\
  resolution factor ($R_2$) plane. {\em Bottom:} The statistical
  uncertainty in the per-bin $a$ estimates.\label{fig:a-2d}}
\end{figure}

In Figure~\ref{fig:a-2d}, we show the additive bias as a function of S/N and resolution factor
for the simulations overall.  For reference, the bottom panel shows the statistical
error for each of the points in parameter space on the top panel.  Similarly to what was done for
shape measurement error in Section~\ref{subsec:results-sigmae}, we fit $a(\text{S/N}, R_2)$ to a
parametric model
\begin{equation}\label{eq:amodel}
a \propto (R_2 - R_{2,0}) \left(\frac{\text{S/N}}{20}\right)^Q,
\end{equation}
with the constant of proportionality, $R_{2,0}$, and $Q$ being free parameters.  Then we interpolate
a correction to that power-law based on the difference between the values in Fig.~\ref{fig:a-2d} to
the best-fitting model.  This defines our baseline additive bias correction in the catalog.

For this baseline correction, the additive bias is negative at low $R_2$ and positive at high $R_2$,
with the deviations from zero being significant only for $\text{S/N}<60$ or $R_2<0.4$.  When fitting
the model in Eq.~\ref{eq:amodel}, the
power-law scaling with S/N is $a\propto (\text{S/N})^{-1.07}$, and the cross-over value
$R_{2,0}=0.51$.

Similarly to what was done in the previous subsection with the multiplicative calibration bias, we
check the sensitivity of this $a$ estimate to how the sample is defined. The results of this
analysis are summarized in Table~\ref{tab:a-sensitivity}.
\begin{table}
  \caption{Sensitivity analysis for additive bias correction $a$.  We present the ensemble average
    $a$ for subsamples of the simulated galaxy sample defined below, after imposing our baseline
    calibration corrections.
}\label{tab:a-sensitivity}
\begin{center}
\begin{tabular}{lll}
\hline
Subsample definition & $a$ & $\delta a^2$ \\ \hline
Sample overall & $-0.008\pm 0.001$ & $2\times 10^{-5}$ \\ \hline
PSF FWHM: 1st quartile & $-0.004\pm 0.002$ & $2\times 10^{-5}$  \\
PSF FWHM: 2nd quartile & $0.001\pm 0.002$ & $4\times 10^{-6}$  \\
PSF FWHM: 3rd quartile & $-0.012\pm 0.003$ & $7\times 10^{-5}$ \\
PSF FWHM: 4th quartile & $-0.006\pm 0.003$ & $3.6\times 10^{-5}$ \\ \hline
$i$ mag $<24$ & $-0.010\pm 0.001$ & $2\times 10^{-5}$\\
$i$ mag $<23.5$ & $-0.007\pm 0.001$ & $1.4\times 10^{-5}$ \\ \hline
\end{tabular}
\end{center}
\end{table}

 In
\citet{2018PASJ...70S..25M} below equation~(24), a requirement on the systematic uncertainty in the
value of $a^2$ is given: $\delta (a^2)$ is required to be below $8\times 10^{-4}$ 
in order to avoid
contaminating a cosmic shear analysis in a way that noticably inflates the systematic error budget.
Given the statistical error in the simulations, as shown in Table~\ref{tab:a-sensitivity}, for our
baseline model we can determine $a$ well enough that the statistical error $\delta a^2$ is more than
an order of magnitude below our requirements.  As a consequence, the systematic uncertainties are
more important.


Comparing the four quartiles in PSF FWHM, we see that the variation in $a^2$ from the two most
extreme values is $1.4\times 10^{-4}$, well below our already-conservative requirements.  For this
reason, we need not worry when dividing up the sample in ways that
result in different seeing distributions that the additive bias behavior will vary significantly
from our baseline model.  Similarly, the cuts on magnitude induce variations in $a^2$ that are also
below our requirements.  We conclude that additive biases due to PSF correction methods can be well-controlled below the level
needed with the existing simulations, without significant ambiguity when imposing other cuts on the
galaxy sample.  Other sources of additive bias, such as PSF modeling errors (as quantified in
\citealt{2018PASJ...70S..25M}), are likely to be more important.

\subsection{Selection bias}\label{subsec:results-selection}

Below we separately describe the results of our analysis of the two types of selection bias
described in Section~\ref{subsec:sel}.

\subsubsection{Selection bias due to weights}

\begin{figure*}
\begin{center}
\includegraphics[width=1.2\columnwidth]{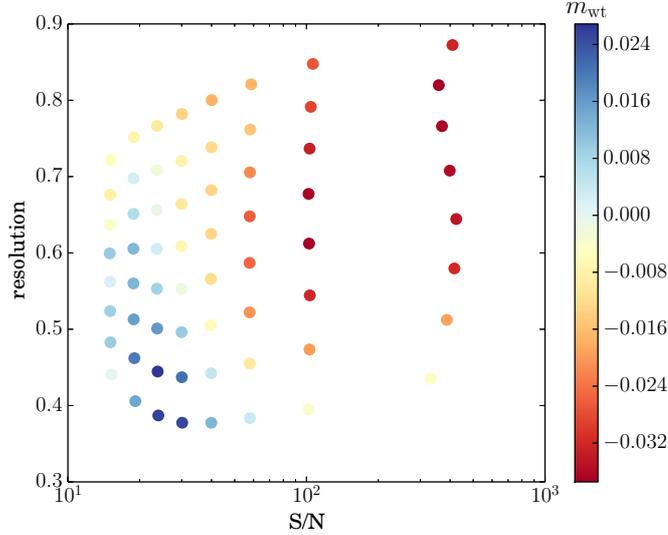}
\end{center}
\caption{The component-averaged shear calibration bias $m_\text{wt}$ due to weight bias in the 2D S/N vs.\
  resolution factor ($R_2$) plane.\label{fig:mwt}}
\end{figure*}

We directly compared the simulation analysis with shape noise cancellation without enforcing the
same per-object weight for each galaxy in a pair, versus the results when we did enforce sameness of
weights.  The resulting multiplicative and additive selection biases are called $m_\text{wt}$ and
$a_\text{wt}$.  We found that $a_\text{wt}$ was consistent with zero within the noise, but there was
a highly statistically significant $m_\text{wt}$ that depends on galaxy properties; see
Figure~\ref{fig:mwt}.  As shown, this bias is negative and reaches a maximum magnitude of $-0.045$
at high S/N and $R_2$, while it is positive and reaches a maximum magnitude of $0.055$ at low S/N
and $R_2$.

We have used the $90^\circ$ rotated pairs to confirm that the sign of the constant of
proportionality in Eq.~\eqref{eq:dx} when $X$ is the lensing weight is consistent with
Fig.~\ref{fig:mwt}.  That is, at low S/N and $R_2$, the lensing weight is larger for galaxies
aligned with the shear than for those oriented perpendicular to it, while at high S/N and $R_2$, it is
smaller for galaxies aligned with the shear.

Since this weight bias depends on the location in this 2D parameter space, we have used the same
process as in Sec.~\ref{subsec:results-shearcalib} to estimate and remove its effects.  The $m$
values given in the catalog include the baseline multiplicative calibration biases due to model and
noise bias, along with $m_\text{wt}$.

\rev{The process described in this section implicitly assumes an analysis of a subset of the data
  with a representative PSF FWHM distribution.  Carrying out a sensitivity analysis for any residual
  biases due to $m_\text{wt}$ varying with the PSF size leads to a conclusion that there are mild
  residuals, of order 0.5 per cent averaged across the whole source sample, when considering extreme
  quartiles in PSF size.  Since we have already concluded in Section~\ref{subsec:results-shearcalib}
  that the baseline calibration bias values $m$ (due to noise and model bias) have an even more
  significant variation for the extreme quartiles in PSF size, and advocated for caution in analyses
  with highly non-representative PSF FWHM distributions, that recommendation will also apply to the
  $m_\text{wt}$ variation with seeing.}


\subsubsection{Selection bias due to cuts}

As described in Section~\ref{subsec:sel}, our formalism involves estimating the multiplicative and
additive selection biases by comparing shear estimation for the sample overall without vs.\ with $90^\circ$
rotated pairs, using that to define the constants of proportionality in Equations~\ref{eq:msel}
and~\ref{eq:asel}, and checking for consistency as we modify the values of the cuts in resolution
factor and magnitude.

For resolution factor, we estimate the shear (including corrections for other forms of shear
calibration bias) with a fixed value for the lower limit in resolution factor, $R_2=0.3$, as we stop
requiring shape noise cancellation around that boundary.  We can check that the formalism described
previously is consistent with the results as we vary $p(R_2=0.3)$ (i.e., $p(\hat{X})_\text{edge}$)
by cutting the sample in other ways and directly inferring the selection bias $m_\text{sel}$.  The
motivation for framing the test in this way -- with a fixed lower limit in resolution factor -- is
that there is no compelling scientific reason to vary the lower limit from $R_2=0.3$.  However, as
we impose cuts on photometric redshifts, the distribution of $R_2$ values gets mildly shifted such
that the relative fraction near the edge of the sample changes at the level of tens of per cent
\citep[see section 8 of][for details of how photometric redshift cuts modify other sample
properties]{2018PASJ...70S..25M}.  Thus the most important thing is to confirm that the
proportionality between $m_\text{sel}$ and $p(R_2=0.3)$ in Eq.~\eqref{eq:msel} holds, and determine
the constant of proportionality.

\begin{figure}
\begin{center}
\includegraphics[width=\columnwidth]{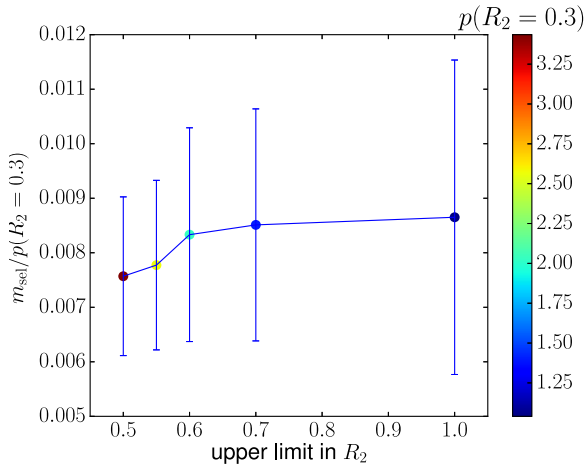}
\end{center}
\caption{This plot shows the empirically-determined (in the simulations) ratio of the multiplicative
  selection bias due to our lower limit in $R_2$ (resolution factor) to the value of $p(R_2)$
  evaluated at that lower limit, which should be a constant according to our
  formalism.  The horizontal axis shows the imposed upper limit in $R_2$ that was used to induce
  variation in $p(R_2=0.3)$ while maintaining shape noise cancellation and hence not causing
  additional selection bias.  The color bar shows the actual value of $p(R_2=0.3)$ for each point. 
Errorbars are correlated between the points due to the use of at least a fraction of the same
simulated source galaxies.
\label{fig:msel-res}}
\end{figure}

Rather than varying photo-$z$ cuts and checking this relationship with the simulations, we consider
more extreme variations in $p(R_2=0.3)$ as a more stringent test of our formalism.  To do so, we
impose strict {\em upper} limits in $R_2$ to remove large portions of the sample, but do so while
imposing shape noise cancellation at that upper edge.  For example, if we impose an upper limit at
the median $R_2$ value, then $p(R_2=0.3)$ should double, and hence the selection bias should as
well.  Our results are summarized in Fig.~\ref{fig:msel-res}.  As shown, the ratio
$m_\text{sel}/p(R_2=0.3)$ is indeed consistent with constant even given a factor of three variation
in $p(R_2=0.3)$, validating our adopted formalism.  For the entire shape sample, the resulting
multiplicative bias is $0.010\pm 0.003$; more generally, if $m_\text{sel}= A \,
p(R_2=0.3)$ then $A=0.0087\pm 0.0026$.  While low-S/N weak lensing measurements such as cluster lensing with individual
clusters can afford to ignore a percent-level bias, more precise galaxy-galaxy or cluster-galaxy
lensing measurements, or cosmic shear measurements, should use this equation and the actual weighted
$p(R_2)$ for their sample with all photometric redshift or other cuts to
correct for the impact of selection bias due to our lower limit in $R_2$.

Using this same approach, we find no statistically significant evidence for additive selection
biases due to either our resolution factor or magnitude cuts, or multiplicative selection biases due
to magnitude cuts.  The errorbars are similar to those for the resolution factor estimate of
selection bias, $\pm 0.003$.  We therefore only impose a correction for multiplicative selection
bias due to the resolution
factor cut.  

\subsection{Dependence of results on parent sample in simulations}\label{subsec:results-samples}

As described in Section~\ref{subsec:galaxymodels}, there are several significant factors in
selection of the parent sample for the simulations.  In this subsection, we present selected results
from simulations produced using parent samples 1--3, in order to ascertain the sensitivity of our
results to parent sample selection, and learn about the systematic uncertainty in
our results.

Referring back to Table~\ref{tab:parentsamp}, we hope to learn the following things by making
certain well-defined comparisons:
\begin{itemize}
\item Comparing the shear calibration biases for simulations with parent samples 1 and 2 will reveal
  the importance of realistic galaxy morphology as compared to simple parametric models.  
\item Comparing results for parent samples 1 and 3 will (at least partially) reveal the impact of
  nearby objects on the outskirts of the primary objects we are measuring.  The reason we are able
  to learn this is that the primary differences between these samples is that in sample 3 we carry out
  deblending only in HSC, without any {\em a priori} deblending in COSMOS (unlike for sample 1).
\item Comparing results for parent samples 3 and 4 will reveal the impact of selecting galaxies in
  {\em HST} based on the cuts given in Section~\ref{subsec:galaxymodels}, rather than selecting them based
  on HSC imaging.
\end{itemize}

\begin{figure*}
\begin{center}
\includegraphics[width=\textwidth]{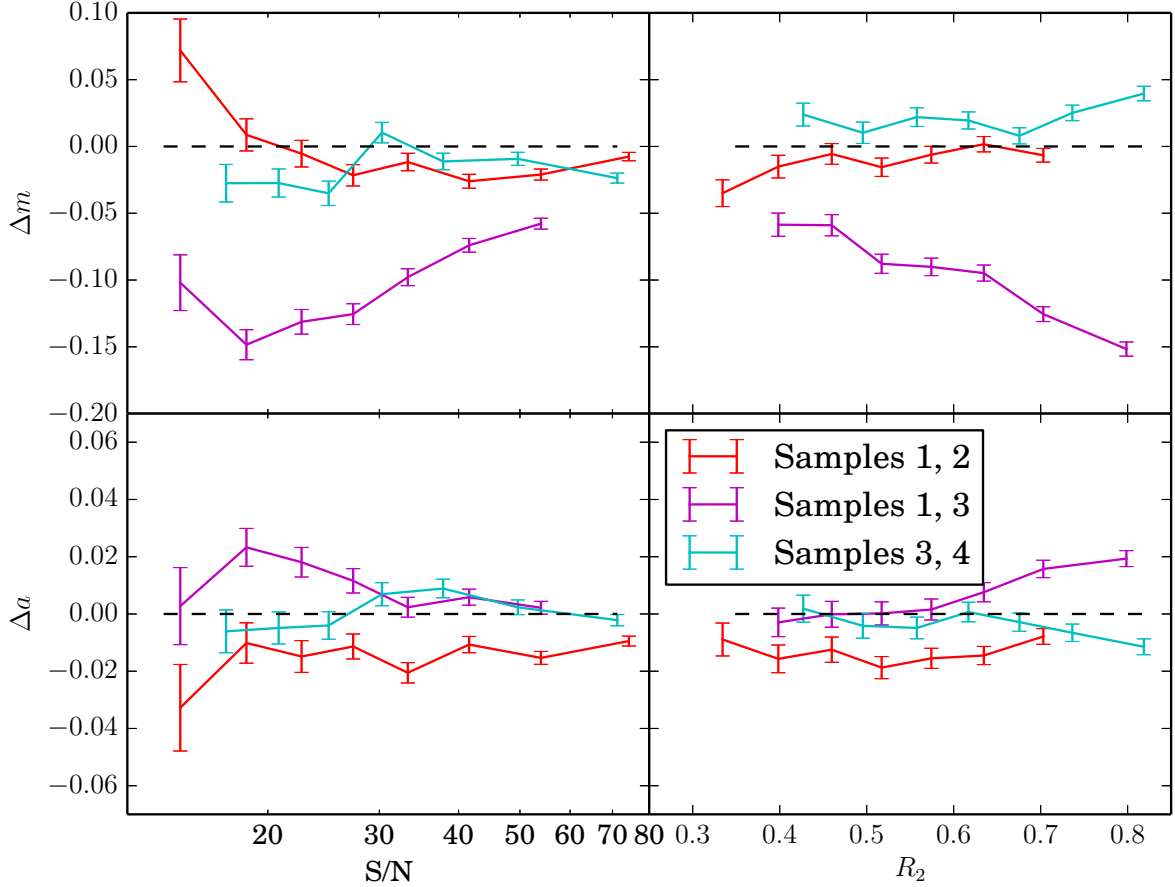}
\end{center}
\caption{Differences in multiplicative bias $\Delta m$ (top) and additive bias $\Delta a$ (bottom) as a function of
  the galaxy S/N (left) and resolution factor (right) when comparing pairs of
  simulations with the different parent samples described in the text.  This figure illustrates the
  importance of the postage stamp creation process in determining the shear calibration.\label{fig:parentsamp_sn}}
\end{figure*}

Figure~\ref{fig:parentsamp_sn} shows differences in calibration bias
$m$ and additive bias $a$ as a function of S/N (left) and resolution factor (right), for these three
pairs of simulation sets.  It is important to bear in mind that when making this comparison in a
single dimension (S/N or $R_2$), we are effectively marginalizing over the distribution of other
galaxy properties.  For example, at a given S/N value, one could in principle get different $m$
values even if the curve of $m(\text{S/N}, R_2)$ is the same for two simulation sets, but the
distribution of $R_2$ values ($\mathrm{d}N/\mathrm{d}R_2$) at that S/N differs for the two simulation
sets.  Hence these plots implicitly include information not just about $m$ and $a$ as a function of
object properties, but also about the joint distribution of object properties. 

\subsubsection{Impact of realistic galaxy morphology}

To consider the impact of realistic galaxy morphology vs.\ parametric models, we consider the curves
for $\Delta m$ and $\Delta a$ on Figure~\ref{fig:parentsamp_sn} 
labeled `Samples 1, 2'. As shown, the $\Delta m$ curve in the top left panel of
Figure~\ref{fig:parentsamp_sn} goes from $\sim 0.07\pm 0.02$ at S/N$\sim$15 to $\sim -0.02\pm
0.005$ at S/N$\sim$80.  The crossing point at S/N$\sim$20 corresponds to the point at which our
distribution of S/N values is cutting off (see Fig.~\ref{fig:data-sims-overall}).  As a result, the
average $\Delta m$ integrated over the whole sample is $-0.014\pm 0.003$.  This is consistent with
the results for many shear estimation methods in the GREAT3 challenge, where
parametric models vs.\ realistic galaxy morphology resulted in differences in shear calibration of
$\lesssim 1$ per cent \citep{2015MNRAS.450.2963M}.  Hence realistic morphology is 
subdominant to effects like noise bias (which results in variations in shear calibration that are an
order of magnitude larger, as shown in Fig.~\ref{fig:m-2d}).  

The bottom left panel of that figure suggests that realistic galaxy morphology can result in differences
in $a$ that are of order 0.01--0.02, consistently across all S/N values.  The sign of this effect is
such that the additive bias is consistent with zero for parametric models, but nonzero for realistic
galaxy morphology.  This suggests that it is harder for re-Gaussianization to remove the PSF
anisotropy from the galaxy shapes when the galaxy models deviate more strongly from simple
elliptical models, and hence is a signature of model bias.

\subsubsection{Impact of nearby objects}

To consider the impact of nearby objects in the COSMOS images, we consider the curves
for $\Delta m$ and $\Delta a$ on Figure~\ref{fig:parentsamp_sn} 
labeled `Samples 1, 3'. As summarized in Table~\ref{tab:parentsamp}, the key difference between
these samples is that for sample 1, we used postage stamps that 
were masked based on using the Sextractor deblender to identify
nearby objects in COSMOS and mask them out with correlated noise.  In contrast, for sample 3, the
nearby objects were not masked, and we ran the HSC object selection and deblending routines.  We
know that if structure is near enough to the central galaxies we are analyzing, then it can be
easily deblended in COSMOS but cannot possibly be recognized as distinct structures in any
ground-based imaging, even with the excellent typical seeing in the HSC $i$-band Wide layer images
(median PSF FWHM of 0.58\arcsec\ for the shear catalog, \citealt{2018PASJ...70S..25M}).  Sample 3
simulations will include both recognized and unrecognized blends, while sample 1 will not.

Clearly $\Delta m$ for these two cases is very large, ranging from $-0.1$ (more negative
bias for sample 3) at S/N$\sim$20 to $-0.03$ at S/N$\sim 80$.  Alternatively, considering the
variation with resolution factor, the difference ranges from -0.07 near our resolution factor limit
$R_2\sim 0.3$, to -0.15 at $R_2\sim 1$.  We note that the galaxies contributing to the two curves at a given S/N value are not
necessarily the same in the two simulation sets, so this can be a combination of multiple effects:
biases at fixed S/N and different biases for objects that have been scattered coherently across S/N
bins.  Whatever the origin of the effect, it is clearly the most striking finding to come from comparing
simulations with different parent samples.  
Since this difference is many times our requirement on systematic uncertainty in $m$ of 0.017, we need to definitively
understand the origin of this difference and ascertain which sample more accurately reflects
reality.  

\begin{figure}
\begin{center}
\includegraphics[width=\columnwidth]{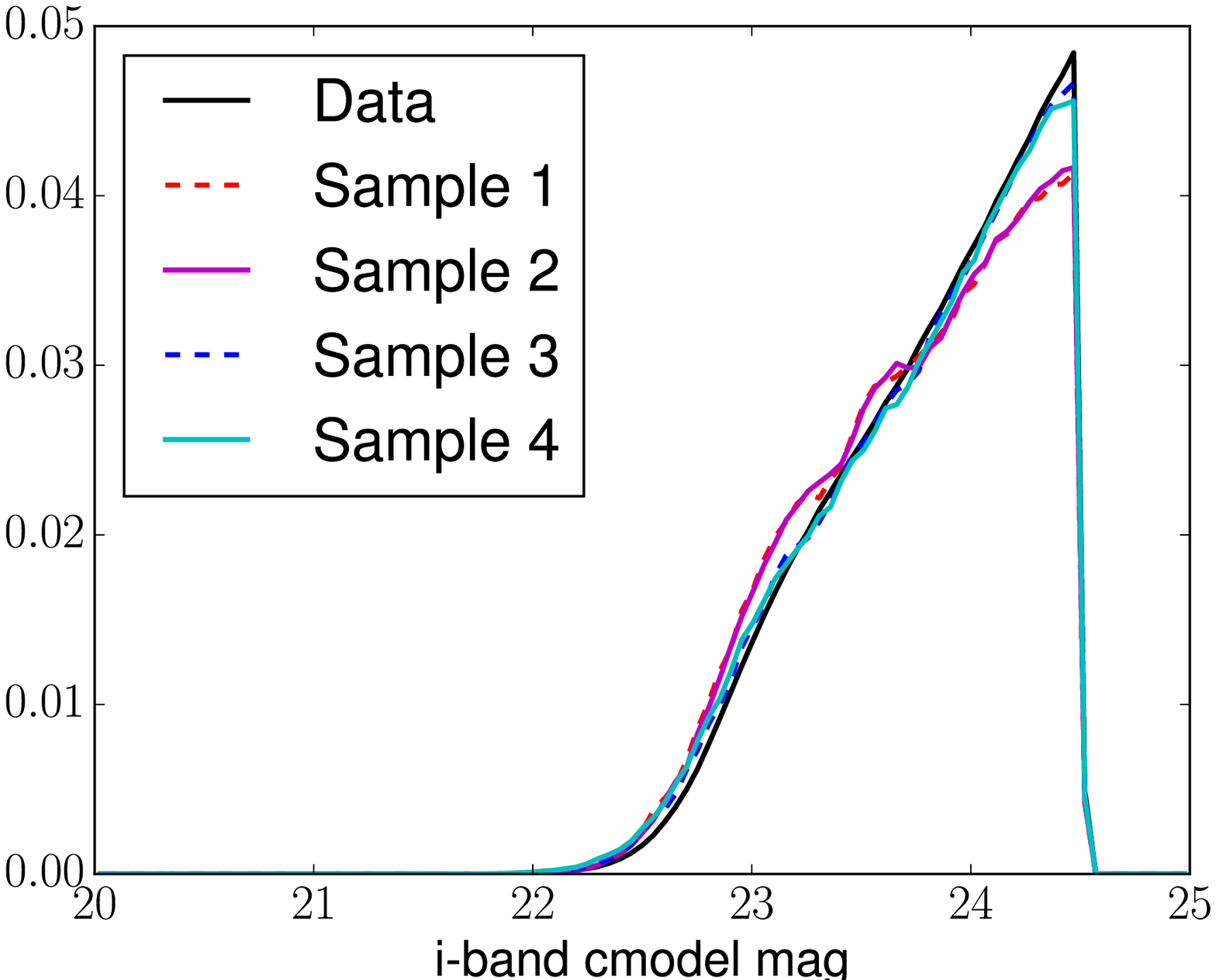}
\includegraphics[width=\columnwidth]{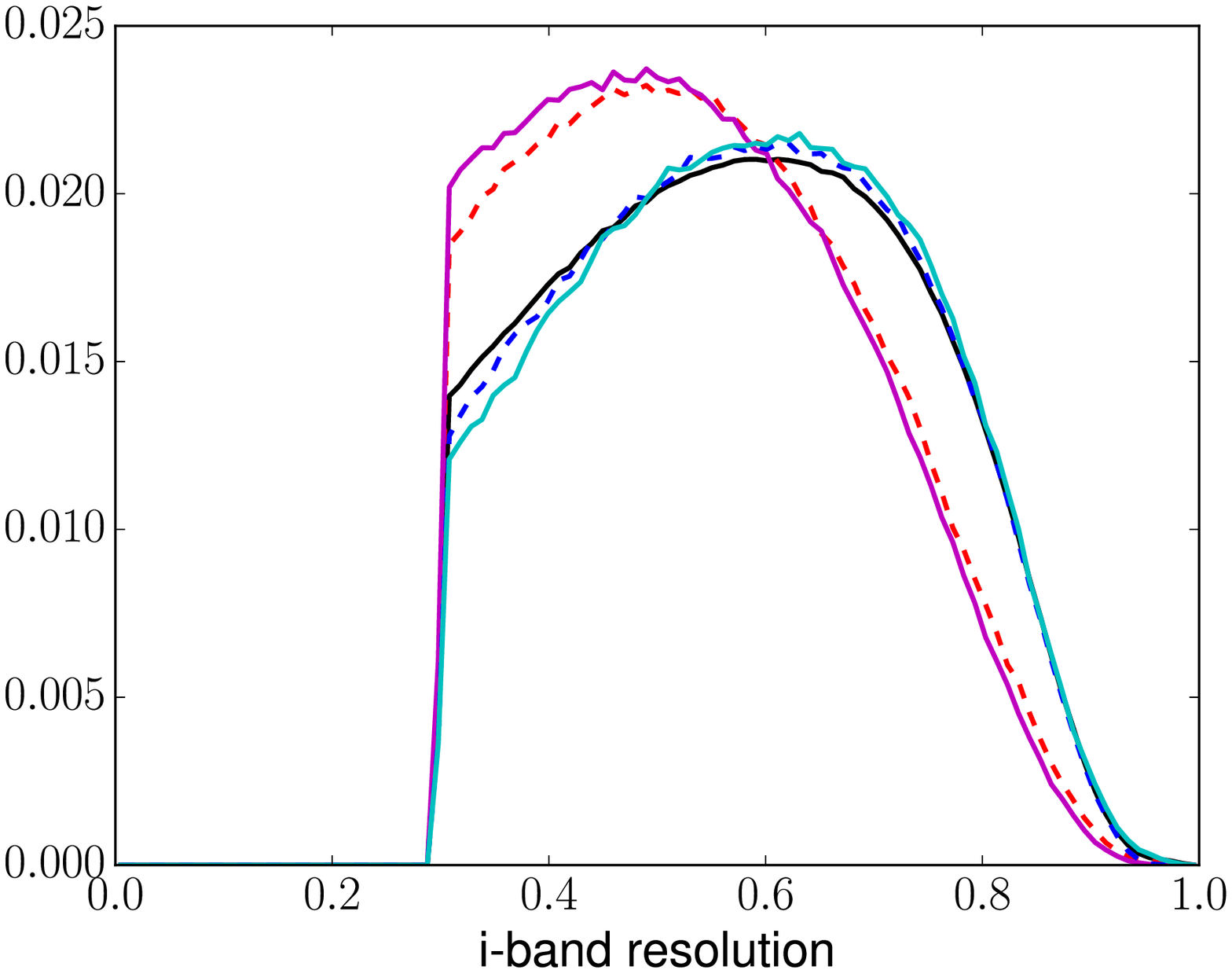}
\end{center}
\caption{Normalized distribution of $i$-band magnitude (top) and resolution factor $R_2$ (bottom) for all
  galaxies passing the lensing cuts in the data and the simulations with the four different parent
  samples, as shown in the legend.  Once we use our procedure defined in this work to
  define the galaxy postage stamps without masking of nearby galaxies (samples 3 and 4), the
  simulated galaxy properties agree well with those in the data.\label{fig:parentsamp_dist}}
\end{figure}

A hint at the resolution of this problem comes from Figure~\ref{fig:parentsamp_dist}, which shows
the distribution of the $i$-band magnitudes (top) and resolution factors (bottom) for galaxies 
passing the weak lensing cuts in the data and in all four parent samples.  The data and parent
sample 4 curves on this plot are the same as those that were shown in
Fig.~\ref{fig:data-sims-overall}.  As shown, the simulations generated using parent samples 1 and 2
(which is similar to sample 1 in that the
masking of nearby objects was also carried out) have distributions of magnitude and resolution
factor that are strikingly different from those in both real data and in the simulations generated
using parent samples 3 and 4 (neither of which has nearby objects masked out in the original COSMOS
images).

The histograms in Figure~\ref{fig:parentsamp_dist} appear to be telling us 
that the galaxies in the simulations made with parent sample 1 are typically
  brighter and smaller than those in real data.  Since parent samples 3 and 4 have nearby
objects in COSMOS that are not masked, it seems clear why we might measure galaxies as being
typically larger in simulations made with those parent samples compared to parent samples 1 and 2:
the extra light on the outskirts of galaxies may inflate the measured galaxy sizes and
resolution factors.  Hence the differences in resolution factor are physically understandable, and
the fact that samples 3 and 4 match the data so much better than samples 1 and 2 in this regard
strongly suggests that the simulations with samples 3 and 4 have captured a physical effect that is
important in determining the apparent resolution factor distribution in real data, due to either the
inability to distinguish blend systems that can be recognized in {\em HST} data or biased resolution
factors for recognized blend systems.  As for the
difference in magnitude distribution, since these are normalized histograms, they must integrate to
the same value.  However, on average, inclusion of light from nearby objects into the central object
light profile is going to make objects appear brighter in the simulations made with parent samples 3
and 4, scattering them into the sample, whereas with samples 1 and 2 they would have failed our weak
lensing cuts.  This explains why parent samples 3 and 4 have relatively more objects that are near
our adopted faint magnitude limit.

The above arguments are valid for two possible explanations of what is going on: (1) as hypothesized
above, light from nearby objects that cannot possibly be deblended in HSC images is artificially
inflating the sizes and magnitudes of some objects, or (2) that the masking process artificially
shredded some low surface-brightness galaxies in the COSMOS images in parent samples 1 and 2, and
thus the masking process was altogether incorrect.  Visual inspection suggests that for objects
brighter than 24th magnitude, process (1) dominates, while near the limiting magnitude, it is
different to discriminate between these two options, and likely both are occurring at some level.
However, regardless of the explanation, Figure~\ref{fig:parentsamp_dist} tells us that if we want
the simulations to accurately represent all relevant image processing effects in the data, we need
to use parent samples 3 and 4, and carry out full object detection and deblending.

While we have shown that the galaxies in the simulations made with parent sample 3 are
systematically larger than those made with parent sample 1, those distributions do not fully explain
the large difference in shear calibration seen in the two simulation sets. 
 To do so, we need to show that for the {\em same galaxies},
the resolution factor is larger in simulations made with parent sample 3 rather than sample 1 (i.e.,
that the difference in $R_2$ distributions cannot be fully explained by new objects scattering in across
the boundary of our cuts).  Since we used the same sample of objects from COSMOS, simply swapping in
unmasked postage stamps for the masked ones with the same simulated PSFs, it is trivial to identify
objects in the two sets of simulations that came from the same object in COSMOS.  However, we
can do better than that: by comparing the masked and unmasked postage stamps, it should be possible
to define a statistic quantifying how much fractional contamination of the object light profile we
expect in the HSC simulations.  Conceptually, our test statistic \rev{$f_\text{contam}$} is defined as the ratio of two
quantities: the weighted sum of flux in pixels that are masked out for sample 1, divided by the
weighted sum of flux in all pixels in the unmasked postage stamps.  The weight function is defined
by convolving the original COSMOS light profile by a Kolmogorov approximation to the simulated HSC
PSF, since we are trying to quantify how much the nearby light would affect the light profile as
seen in HSC.  If our explanation is correct, then the difference in resolution factor
and/or magnitude $\Delta R_2$ and $\Delta$mag between the simulation sets should be a strong
function of \rev{$f_\text{contam}$}, nearing zero for apparently isolated objects.  

\rev{To identify a statistical correlation between $f_\text{contam}$ and $\Delta R_2$
  and/or $\Delta$mag, we do not calculate a correlation coefficient, because we do not necessarily
  expect a linear relationship.  Instead, we divide the sample based on an $f_\text{contam}$ value
  corresponding to low (few per cent) contamination of the light profile vs.\ higher contamination,
  and compare the typical $\Delta R_2$ and $\Delta$mag values for objects above and below that level
of contamination.  This dividing line happens to fall around the 70th percentile in
$f_\text{contam}$.
We find that if we split the sample into those above vs.\ below the 70th percentile in $f_\text{contam}$}, the $\Delta R_2$ is typically 0.07 vs.\ 0.01.  That bias in resolution factor should give
rise to a difference in shear calibration that is of order 0.06 (with the exact value depending on
the entire distribution of object properties).  It is therefore plausible that between this bias in
$R_2$ and the fact that new objects are scattered into the sample by the light from nearby objects,
changing the galaxy population overall, the shear calibration bias differences seen in simulations
with samples 1 and 3 can be entirely explained.  Similar, $\Delta$mag for those above vs.\ below the
70th percentile in our test statistic is $-0.15$ vs.\ $-0.02$, supporting our argument that some
galaxies from below the flux limit in the simulations with parent sample 1 are getting scattered
across our cuts and into the weak lensing-selected sample in parent sample 3 (and the real data).

By splitting the galaxy population in the sample 3 simulations at
the 70th percentile in our test statistic, we have also directly ascertained  that the shear calibration biases for the two
subpopulations within that simulation sets differ by approximately the amount expected based on this
bias in the $R_2$ values.
%

This dramatic effect due to different parent sample populations is related to an effect that has
been reported elsewhere in the literature.  \cite{2015MNRAS.449..685H} carried out tests of shear
estimation using the moments-based KSB method, and argued based on their figure 7 that in order to
properly estimate the shear calibration for a sample with some limiting magnitude $m_\text{lim}$, it
is necessary to include galaxies about 1.5 magnitudes fainter than $m_\text{lim}$ in the
simulations.  Their explanation was that these galaxies contribute a small amount of light that
essentially looks like correlated noise clumps on the outskirts of the galaxies brighter than
$m_\text{lim}$.  It is worth considering that both KSB and re-Gaussianization are moments-based
methods that use weighted moments to correct the observed shapes for the blurring by the PSF.  If
the nearby objects {\em systematically} cause a positive bias in the observed size 
of the galaxies, then the estimated shears will systematically be too low by a similar factor for
both methods. \rev{It is less obvious whether this effect will cause a similar problem for
  model-fitting methods, and likely depends on implementation details of the method.} 
However, there is an important difference between what was
reported in \cite{2015MNRAS.449..685H} and what is shown here.  They report that the primary cause of the effect
they see is due to unresolved backgrounds contributing correlated noise.  In HSC, we have set a quite
conservative magnitude limit, and the objects that are nearby and were masked in parent sample 1
would typically not be below the detection limit in HSC (despite the fact that, on their own, they
would be too faint to pass our conservative weak lensing cuts).  Thus we conclude that in HSC, the
effect we are seeing is primarily due to light from nearby objects above the detection threshold
that lead to the aforementioned bias in resolution factor and shear.  Indeed, unresolved backgrounds
contribute correlated noise near our simulated galaxies in both simulation sets 1 and 3.

 For the Dark Energy Survey Year 1 analysis,
\citet{2018MNRAS.475.4524S} used simulations to study the impact of nearby objects on shear
estimation, and identified a net calibration bias (from several effects combined) that is of order
30 to 100 per cent of the effect seen in this work, depending on the sample selection.  They
find that removal of nearby objects is an effective mitigation scheme, while for HSC we do not
expect this approach to be effective because our depth results in a higher fraction of unrecognized
blends, despite the smaller seeing size.  As noted in that work, it seems unlikely that this effect
can be represented as purely a scale-independent calibration bias; however, for the level of precision
in shear correlation functions in the first-year HSC analysis, this approximation is likely to be sufficient.

Returning to the trends in $\Delta m$ between simulations made with samples 1 and 3 in
Figure~\ref{fig:parentsamp_sn}, we found more severely negative
calibration biases at low S/N and high $R_2$.  This makes sense given the origin of the effect.  Low
S/N objects will have a higher fractional contamination of their light profile by nearby objects,
resulting in a larger bias in $R_2$ and the shear.  Similarly, objects with larger size may have
been artificially inflated by this effect, or may be truly large objects with low surface
brightness, in either case resulting in a larger shear bias. 

\begin{figure}
\begin{center}
\includegraphics[width=\columnwidth]{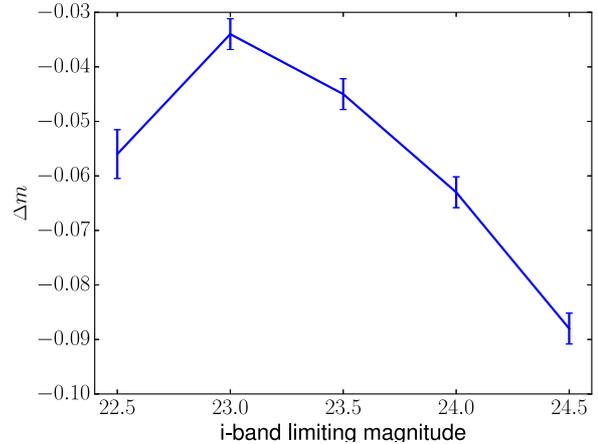}
\end{center}
\caption{Additional calibration bias due to the presence of nearby objects in the galaxy postage stamps, as a
  function of the limiting magnitude of the sample selected for shear estimation.\label{fig:bb_mag}}
\end{figure}

In principle, the size of this shear bias should depend on the number density of sources of
sufficient flux to cause a measurable bias in the shape of the galaxies of interest.  Assuming only
galaxies within a fixed $\Delta$mag of the limiting magnitude can have an effect, and knowing that
the number density of brighter objects is lower than that of fainter objects, we expect a smaller
effect for samples with a brighter limiting magnitude.  Using the simulations, we can directly
measure this trend by changing the limiting magnitude from 24.5 to something brighter than that, and
comparing parent samples 1 and 3.  The results for the average $\Delta m$ (over the entire samples
brighter than some limiting mag) are shown in Fig.~\ref{fig:bb_mag}.  The $\Delta a$ (not shown on
the plot) is in all cases consistent with zero.  As shown, the shape of the curve is largely what we
expect; however, there is a slight turnover at the brightest limiting magnitude that is difficult to
explain.  Unfortunately we cannot test any further restriction on the sample, as the number of
galaxies in the simulations brighter than $i=22$ is too small.

Finally, we note that another difference between samples 1 and 3 shown in Table~\ref{tab:parentsamp}
is the imposition of the `marginal' selection criteria on the simulations made with sample 1.  We
have explicitly tested whether this can be responsible for some of the differences seen, by imposing
the `marginal' cut on the simulations with parent sample 3 after their creation, and found that the
effect of this cut on the shear calibration in the simulations is negligible.

\subsubsection{Impact of HSC vs.\ {\em HST} selection}

To consider the impact of selecting the galaxies with the flag cuts described in
Section~\ref{subsubsec:deblps} in {\em HST}, versus selecting them based on their detectability in HSC as
described in Section~\ref{subsubsec:finalps}, we consider the curves for $\Delta m$ and $\Delta a$
on Figure~\ref{fig:parentsamp_sn} labeled `Samples 3, 4'. As
shown, the $\Delta m$ curves suggest that different choices of how to select the galaxies can induce
$\sim 2$ per cent differences in shear calibration.  However, as shown in
Figure~\ref{fig:parentsamp_dist}, the resulting differences in the selected object properties are
quite small.

In this case, none of our diagnostics revealed obvious flaws with the sample 3 simulations.
Nonetheless, as described in Section~\ref{subsubsec:deblps}, diagnostics of the parent samples
indicated that the ground-based selection flag applied to the COSMOS sample preferentially removes
galaxies that are near other bright galaxies from the sample.  Hence, as a matter of principle, we
adopt sample 4 as the fiducial one.

\section{Conclusions}\label{sec:conclusions}

The over-arching goal of this paper is to use simulations with a realistic galaxy population, and HSC
imaging conditions, to constrain certain weak lensing shear systematics that are difficult to
identify using the data itself, devise corrections for our best estimates of those systematics, and
estimate systematic uncertainties in those corrections in order to build our systematic error budget
for cosmological weak lensing analyses with the HSC weak lensing shear catalogue described in
\citet{2018PASJ...70S..25M}.  For this purpose, we have used large suites of simulations with galaxy
images directly from the {\em HST} COSMOS imaging, selected based on deep coadded HSC imaging, with
the {\em HST} PSF removed, lensing shear applied, and then applied further image manipulations to look like
the HSC survey data.  While the analysis used postage stamp images, no attempt was made to mask out
nearby objects from the outset, and the selection and deblending parts of the HSC pipeline were run.
Hence the impact of nearby objects and unrecognized blends (blends of two or more objects that are not
recognized as such by the HSC pipeline) is implicitly included, albeit with the assumption that the
two objects have the same shear.

More specifically, the tests in this paper were focused on understanding multiplicative and
additive shear biases due to the following effects:
\begin{enumerate}
\item Model bias due to fully realistic galaxy morphology (no parametric models) and other
  fundamental failures of the assumptions of the re-Gaussianization shear estimation method;
\item noise bias;
\item the impact of light from nearby objects that are properly detected and deblended;
\item unrecognized blends that are at the same or similar redshifts;
\item selection bias due to the weight factors used for ensemble averages depending on the galaxy
  shape (`weight bias'); and
\item selection bias due to the quantities used to place cuts on the sample depending on the galaxy
  shape.
\end{enumerate}

This list is by no means an exhaustive coverage of the observational systematic error budget for a
cosmological weak lensing analysis.  There are three types of effects that were excluded.  The first
type of effects were excluded because it is not possible to quantify them with the type of
single-band simulations made here: photometric redshift errors; the impact of unrecognized blends not
at the same redshift (and not carrying the same shear) for both shear estimation and photometric
redshifts; and selection biases in shear induced by photometric redshift cuts.  
The second type of effects were excluded because they are
constrained elsewhere \citep{2018PASJ...70S..25M}:  PSF modeling errors, relative astrometric
errors, and stellar contamination.  There is a final class of effects that were
not covered here because we anticipate they are too low-level to form a significant part of the HSC
first-year weak lensing error budget: e.g., the impact of chromatic PSFs or non-weak shear.

In evaluating the quality of the simulations themselves, and the fidelity with which they represent
the data, we identified an important issue with the input galaxy population.  Based on our findings,
the process of masking out nearby objects in the original {\em HST} images makes it impossible to
simulate a realistic ground-based survey, because the non-negligible fraction of unrecognized blends
that will not be accurately represented if nearby objects in {\em HST} images are masked.  The
simulated galaxy populations appear to be too small on average, and have the wrong magnitude
distribution, if the masked images are used.  This was found to be an important factor in shear
estimation, which we have interpreted as a sensitivity of the re-Gaussianization algorithm to both
light from nearby objects and unrecognized blends.  Hence, the results that we use for science are
the ones that include the impact of both recognized and unrecognized blends.

We have quantified and removed non-negligible systematic biases in the ensemble shear estimates due
to the six sources of bias enumerated earlier in this section.  In
Sections~\ref{subsec:results-shearcalib} and~\ref{subsec:results-add}, we have quoted requirements
on the uncertainties in multiplicative and additive bias from \citet{2018PASJ...70S..25M}, and discussed how well we meet
those requirements for the specific set of systematics addressed in this paper. We found that the
additive biases were not strongly sensitive to divisions of the source sample based on seeing size
or magnitude, and hence cuts on the source sample for a real analysis should not cause a failure to
meet our requirements.  The multiplicative bias corrections defined here were found to have a modest dependence
on seeing size, such that if we use a special sample of lenses that only resides in a particular
region of the survey that has anomalously good or bad seeing, it may be necessary to redefine the
calibration factors using the simulations.  However, our core cosmology analyses that use the entire
first-year area set our most stringent requirements, and are not subject to this limitation.
Similarly, cuts in magnitude (which was not one of the parameters used to define our calibration
corrections) were found to cause marginally statistically significant ($2.5\sigma$) multiplicative
biases, but these were still $3\times$ lower than the conservative requirements set in
\citet{2018PASJ...70S..25M}, and hence we do not anticipate failures of our multiplicative bias
corrections for magnitude- or color-selected samples.


Overall, our findings suggest that the quantifiable biases due to the causes investigated in this
paper are significant in magnitude, but understood and therefore corrected at the level of precision
required for first-year HSC weak lensing analysis (to within roughly 1 per cent).  In terms of the future outlook, having an
independently-developed shear estimation method with completely different assumptions will be an
important sanity check on the results in this paper; forthcoming results from
Armstrong et~al.~({\em in prep.}) should provide this sanity check.  In addition, one of the key results of
this paper -- the sensitivity of the results to the simulated galaxy population -- suggests that for
future shear catalogs in HSC covering larger areas and requiring better systematics control, we may
wish to adopt shear calibration methods that do not require the ground-up generation of realistic
image simulations, such as the metacalibration method
\citep{2017arXiv170202600H,2017ApJ...841...24S}.  A further benefit of this change is that it would
enable us to use galaxy samples fainter than $i=24.5$, which would otherwise require deeper {\em
  HST} training data.

\section*{Acknowledgments}

We thank Simon Samuroff \rev{and the anonymous referee} for helpful comments on this work.  
RM and FL acknowledge the support of the US Department of Energy Early Career Award Program.
Support for this work was provided by the University of California 
Riverside Office of Research and Economic Development through the
FIELDS NASA-MIRO program. A portion of this research was carried out at the Jet Propulsion
Laboratory, California Institute of Technology, under a contract with
the National Aeronautics and Space Administration.

The Hyper Suprime-Cam (HSC) collaboration includes the astronomical communities of Japan and Taiwan,
and Princeton University. The HSC instrumentation and software were developed by the National
Astronomical Observatory of Japan (NAOJ), the Kavli Institute for the Physics and Mathematics of the
Universe (Kavli IPMU), the University of Tokyo, the High Energy Accelerator Research Organization
(KEK), the Academia Sinica Institute for Astronomy and Astrophysics in Taiwan (ASIAA), and Princeton
University. Funding was contributed by the FIRST program from Japanese Cabinet Office, the Ministry
of Education, Culture, Sports, Science and Technology (MEXT), the Japan Society for the Promotion of
Science (JSPS), Japan Science and Technology Agency (JST), the Toray Science Foundation, NAOJ, Kavli
IPMU, KEK, ASIAA, and Princeton University.

This paper makes use of software developed for the Large Synoptic Survey Telescope. We thank the
LSST Project for making their code available as free software at  \url{http://dm.lsst.org}.

The Pan-STARRS1 Surveys (PS1) have been made possible through contributions of the Institute for
Astronomy, the University of Hawaii, the Pan-STARRS Project Office, the Max-Planck Society and its
participating institutes, the Max Planck Institute for Astronomy, Heidelberg and the Max Planck
Institute for Extraterrestrial Physics, Garching, The Johns Hopkins University, Durham University,
the University of Edinburgh, Queen's University Belfast, the Harvard-Smithsonian Center for
Astrophysics, the Las Cumbres Observatory Global Telescope Network Incorporated, the National
Central University of Taiwan, the Space Telescope Science Institute, the National Aeronautics and
Space Administration under Grant No.\ NNX08AR22G issued through the Planetary Science Division of the
NASA Science Mission Directorate, the National Science Foundation under Grant No. AST-1238877, the
University of Maryland, and Eotvos Lorand University (ELTE) and the Los Alamos National Laboratory.

Based on data collected at the Subaru Telescope and retrieved from the HSC data archive system,
which is operated by Subaru Telescope and Astronomy Data Center, National Astronomical Observatory
of Japan.


\bibliographystyle{mnras}
\bibliography{../../catalog_paper/papers}

\appendix

%

\section{Shape measurement error formalism}\label{app:sigmae}

This appendix outlines the formalism for using $90^\circ$ rotated pairs of galaxies in the
simulations to estimate the shape measurement error as a function of galaxy properties. \rev{The
  formalism below assumes the shape definition in Equation~\eqref{eq:e}.}

For each such pair, after analyzing the simulations with the HSC pipeline, we have an estimate of
the distortion for both galaxies, $\hat{e}$ and $\hat{e}_\text{rot}$ (consider these as
two-component complex ellipticities).  These are related to the true intrinsic shape $e_\text{int}$
(or $-e_\text{int}$ for the 90$^\circ$ rotated version), the lensing reduced shear $g$, and the
independent measurement errors for the two realizations $\delta e$ and $\delta e_\text{rot}$ as:
\begin{eqnarray}
\hat{e} &= \frac{e_\text{int} + 2g + g^2 e_\text{int}^*}{1 + |g|^2 + g e_\text{int}^* + g^* e_\text{int}} + \delta e\\
\hat{e}_\text{rot} &= \frac{-e_\text{int} + 2g - g^2 e_\text{int}^*}{1 + |g|^2 - g e_\text{int}^* -
  g^* e_\text{int}}  + \delta e_\text{rot}.
\end{eqnarray}

In the weak shear limit ($g^2\approx |g|^2\approx 0$), these equation simplify to
\begin{eqnarray}
\hat{e} &= \frac{e_\text{int} + 2g}{1 + g e_\text{int}^* + g^* e_\text{int}} + \delta e\\
\hat{e}_\text{rot} &= \frac{-e_\text{int} + 2g}{1 - g e_\text{int}^* - g^* e_\text{int}}  + \delta e_\text{rot}
\end{eqnarray}
or, even further, to
\begin{eqnarray}
\hat{e} &= (e_\text{int} + 2g)(1 - g e_\text{int}^* - g^* e_\text{int}) +\delta e\\ &\approx e_\text{int} + 2g -g
|e_\text{int}|^2 - g^* e_\text{int}^2+ \delta e\\
\hat{e}_\text{rot} &\approx -e_\text{int} + 2g -g
|e_\text{int}|^2- g^* e_\text{int}^2 + \delta e_\text{rot}.
\end{eqnarray}

Now we consider what we can learn from the estimated shapes for our $90^\circ$-rotated pairs by
taking the sum of $\hat{e}$ and $\hat{e}_\text{rot}$:
\begin{equation}
\hat{e}+\hat{e}_\text{rot} = 4g  - 2 g |e_\text{int}|^2 - 2 g^* e_\text{int}^2 + \delta e + \delta e_\text{rot}
\end{equation}
Note that we have not taken ensemble averages and hence no terms have been cancelled due to the
intrinsic random orientations of galaxies.

However, if we consider the first term on the right compared to the second and third, we know that
the typical intrinsic distortion means the first exceeds the second and third by nearly an order of
magnitude.  In addition, in the simulations we precisely know the input shear $g$, and hence we can
move this to the left-hand side (folding it into our estimated quantities).  Hence our final equation is
\begin{equation}
\hat{e}+\hat{e}_\text{rot} - 4g \approx \delta e + \delta e_\text{rot} \equiv \hat{e}_\text{sum}
\end{equation}

The new quantity that we have constructed, $\hat{e}_\text{sum}$, is distributed like the sum of two
random draws from the shape measurement error distribution $p_\text{err}(e)$. The probability
distribution for the sum $\delta e + \delta e_\text{rot}$ is the convolution of $p_\text{err}(e)$
with itself.  Without assuming the form of $p_\text{err}(e)$ (e.g., we do not want to assume it is
Gaussian), we can at least infer the variance of the distribution\footnote{Technically we are
  assuming one thing about $p_\text{err}(e)$: that it has well-defined second moments (unlike, for
  example, 
  a Cauchy distribution), but this is a relatively weak assumption.}, by taking the variance
of the $\hat{e}_\text{sum}$ values and halving it.

\end{document}